\documentclass[onecolumn]{emulateapj}    
\usepackage{eqsecnum}

\newcommand{\gtapprox}{\raisebox{-0.5ex}{$\,\stackrel{>}{\scriptstyle
\sim}\,$}}
\newcommand{\ltapprox}{\raisebox{-0.5ex}{$\,\stackrel{<}{\scriptstyle
\sim}\,$}}
\newcommand{\ro}{r_{\rm i}}
\newcommand{\rg}{r_{\rm g}}

\newcommand{\vK}{v_{\scriptscriptstyle \rm K}}
\newcommand{\vA}{v_{\scriptscriptstyle \rm A}}
\newcommand{\cs}{c_{\rm s}}
\newcommand{\phiG}{\phi_{\scriptscriptstyle \rm G}}
\newcommand{\OmegaK}{\Omega_{\scriptscriptstyle \rm K}}
\newcommand{\zetaK}{\zeta_{\scriptscriptstyle \rm K}}

\newcommand{\uB}{u_{\scriptscriptstyle \rm B}}
\newcommand{\uK}{u_{\scriptscriptstyle \rm K}}
\newcommand{\tR}{t^{\scriptscriptstyle \rm R}}
\newcommand{\tB}{t^{\scriptscriptstyle \rm B}}
\newcommand{\tv}{t^{\rm v}}
\newcommand{\FE}{F^{\scriptscriptstyle \rm E}}
\newcommand{\Ld}{L_{\rm d}}

\newcommand{\Pa}{P_{\rm a}}

\newcommand{\PQ}{P_{\scriptscriptstyle \rm Q}}

\newcommand{\Pw}{P_{\rm w}}
\newcommand{\sigmaT}{\sigma_{\scriptscriptstyle \rm T}}

\newcommand{\tauT}{\tau_{\scriptscriptstyle \rm T}}

\newcommand{\Mdot}{\dot M}
\newcommand{\Mdota}{\dot M_{\rm a}}
\newcommand{\Mdotw}{\dot M_{\rm w}}

\newcommand{\dr}{\frac{\rm d}{{\rm d}r}}
\newcommand{\dz}{{\rm d}z}

\received{}
\revised{}
\accepted{}

\shortauthors{Kuncic \& Bicknell}
\shorttitle{MHD Accretion in AGN}

\begin{document}

\title{Dynamics and Energetics of Turbulent, Magnetized Disk Accretion
around Black Holes:
a First-Principles Approach to Disk--Corona--Outflow Coupling}
\author{Zdenka Kuncic\altaffilmark{1}}
\affil{School of Physics, University of Sydney}
\email{Z.Kuncic@physics.usyd.edu.au}
\and
\author{Geoffrey V. Bicknell\altaffilmark{2}}
\affil{Research School of Astronomy \&
Astrophysics \\
Australian National University}
\email{Geoff.Bicknell@anu.edu.au}
\altaffiltext{1}{Postal address: School of Physics, University of Sydney, Sydney, NSW 2006, Australia}
\altaffiltext{2}{Postal address: Mt Stromlo Observatory, Cotter Rd., Weston, ACT 2611, Australia.}

\begin{abstract}
We present an analytic description of turbulent,
magnetohydrodynamic (MHD) disk accretion around black holes that
specifically addresses the relationship between radial and vertical, mean-field transport of mass,
momentum and energy, thereby complementing and extending
numerical simulations. The azimuthal-vertical component of the magnetic stress is fundamental to an understanding of disk--corona--outflow coupling: when it is important for driving the angular momentum transport and mass accretion in the disk, it also has an important influence on the disk--corona--outflow energy budget. The Poynting flux derived from the product of this term with the Keplerian velocity also dominates the Poynting flux into the corona. The ratio of the coronal Alfv\'{e}n velocity to the Keplerian velocity is an important parameter in disk-corona-outflow physics. If this parameter is greater than unity then energetically significant winds and Poynting flux into the corona occur. However, significant effects could also occur when this parameter is much less than unity.
A limiting solution describing the case of angular momentum transport solely by the vertical-azimuthal stress has the property that all of the accretion power is channeled into a wind, some of which would be dissipated in the corona. More realistic solutions in which there is both radial and vertical transport of angular momentum would have different fractions of the accretion power emitted by the disk and corona respectively. These results have important implications for existing accretion disk
theory and for our interpretation of high-energy emission and
nuclear outflows from the central engines of Active Galactic Nuclei and Galactic Black Hole Candidates.
\end{abstract}


\section{Introduction}

Since the foundations were laid for a standard theory of disk accretion
\citep{pringrees72,SS73,novthorn73}, two fundamental problems were immediately 
recognized  \citep{liangprice77,bisblin77,pacz78}:
(1) The efficient transport of angular momentum to large radii
cannot be attributed to conventional kinematic viscosity, and (2) The observed
high-energy spectra and luminosities and the ubiquity of outflow phenomena
from accreting black holes implies efficient vertical transport of
energy from a relatively cool, dense disk to a hot, tenuous and 
unbound corona.
While it has been widely accepted that magnetic fields provide the most plausible
means of efficiently transporting both angular momentum and energy, the precise
nature of this transport has remained unclear until recent numerical simulations
demonstrated, unambiguously, that accretion disks work because of magnetohydrodynamic
(MHD) turbulence \citep*{balbhaw98,hawley99}.
The turbulence is generated by the magnetorotational instability
\citep[MRI -- ][ and references therein]{balbhaw91}, which is
driven by the free energy available from the differential rotation of the bulk flow.
Notwithstanding these groundbreaking results, however, the nature of vertical energy
transport from an accretion disk to a corona and/or outflow still remains an outstanding
and contentious issue.

In the context of Active Galactic Nuclei (AGN), much theoretical effort has
recently focussed on the physics of accretion disk coronae
\citep*{dimatt97a,dimatt97b,merlfab01a,merlfab01b,merlfab02,liu02}, commensurate with the
dramatic increase in both the quality and quantity of high-energy observational data.
However, there has been little improvement in coupled disk--corona models since the
first phenomenological descriptions of \citet{haardt91,haardt93}.
Current models \citep[e.g.][]{merlfab02,liu02} simply replace the  fraction of
accretion power transferred from the disk to the corona with a Poynting flux
quantity estimated from a mean-field buoyant velocity and an
equipartition, mean-field magnetic energy density.
While numerical models \citep[e.g.][]{millstone00} do indeed show that turbulent
fluctuations in a vertically stratified disk are capable of driving the
magneto-gravitational modes of the Parker instability \citep{parker55}, whether
magnetic buoyancy can supply the corona with sufficient power to explain the observed
high-energy emission  is questionable.
Numerical simulations indicate that magnetic buoyancy is an ineffective saturation
mechanism for the MRI \citep{brandenburg95,stone96,millstone00}, while theoretical
models for disk coronae require implausibly ideal buoyancy conditions and limiting
accretion conditions  \citep[e.g.][]{merloni03}.
Realistically, the growth of the unstable buoyant-Parker modes, which is essentially
a wave-fluctuation resonance  interaction, must compete against
particle-fluctuation interactions, which correspond to dissipation of
the turbulence and internal heating of the disk.

The production and ubiquity of outflows from accretion disks around
black holes also remains a challenging problem and it is unclear from
most theoretical models
\citep*[e.g.][]{blandpayne82,lovelace91,li92,wardkon93,ustyugova00}
whether MHD disk turbulence plays a significant role
\citep[but see][]{heinzbeg00}.
Nevertheless, numerical simulations of turbulent MHD accretion disks
do in fact show that outflows become important
in the innermost regions of turbulent accretion disks
\citep*[e.g.][]{stonepring01,hawley01,hawbalb02}.
Unfortunately, the numerical models are restricted by their approach:
non-Keplerian motions are defined {\it a priori} as fluctuating quantities,
so that vertical, mean-field transport is not self-consistently taken
into account.
Indeed, the outflows that emerge in some of the simulations \citep[e.g.][]{hawbalb02}
are defined as regions where the nett radial flow is outward, rather than inward.
Furthermore, the  neglect of vertical angular momentum transport restricts
the radial inflow to substantially subsonic speeds \citep[see][]{balbhaw98}.
Inevitably, the resulting accretion rates in the numerical models are typically
very low \citep[see e.g.][]{stonepring01,hawley01,hawbalb02}.
Numerical models are also restricted by computational limitations: simulations that
are global as well as vertically stratified are required to estimate the fraction of
accretion power that can be vertically transported and this is not only computationally
prohibitive, but also sensitive to numerical dissipation effects, which are
difficult to quantify.

Thus, the present status of black hole accretion disk theory is that
{\em there is currently no formalism which self-consistently couples turbulent, MHD
disk accretion with a magnetically-dominant corona in a framework that can
accomodate a range of radial inflow and vertical outflow solutions.}

In this paper, we present the first fully analytic description of a turbulent
MHD accretion disk coupled to a corona, self-consistently taking into account both
vertical and radial mean-field fluxes of mass, momentum and energy.
By deriving the relevant transport equations from first-principles,
our formalism provides a non-phenomenological and non-empirical
approach to the problem of energy transport to a corona and the associated outflow and the inter-relationship of this transport with accretion onto the central black hole.
In this treatment, we focus on the effects of turbulent magnetic
stresses and the mean magnetic flux density is assumed to be zero,
for the sake of clarity.
However, our formalism lends itself naturally to the inclusion of nonzero
mean magnetic fields, which we intend to explore separately in a subsequent paper.
In \S~2, we present the relevant conservation equations for a resistive,
viscous, and optically-thick MHD gas.
In \S~3, we statistically average these equations and derive the
corresponding mean-field equations for a turbulent MHD gas.
In \S~4, we apply these mean-field equations to the dynamics of a
geometrically-thin accretion disk that is stationary and axisymmetric in the mean,
and we expressly examine the implications of vertical mean-field transport for the
conservation of mass and momentum.
In \S~5, we utilize the results of \S~4 to analyze the total disk energy budget.
We conclude with a discussion of the main results in \S~6.

\section{Fundamental equations}
\label{s:fundamental}

In this section, we summarize the standard equations for a resistive, viscous,
and radiative MHD gas in a dynamically important gravitational field.
We consider only a nonrelativistic, optically-thick gas in which the radiative
diffusion approximation holds, so that the radiation pressure reduces to a scalar.
The independent variables are: mass density, $\rho$; fluid velocity, $v_i$;
gas plus radiation pressure, $p$; gas plus radiation energy density, $u$;
gravitational potential, $\phiG$;
radiative flux, $F_i$; external heat flux, $Q_i$;
magnetic field, $B_i$; electric field, $E_i$; current density, $J_i$;
and viscous stress tensor, $\tv_{ij}$\footnote{Note that this is the viscous stress tensor describing microscopic processes. It does {\em not} represent the so-called turbulent viscosity.}

The fundamental equations are:

\begin{enumerate}


\item Mass continuity.
\begin{equation}
\label{e:mass}
\frac{\partial \rho}{\partial t} \, + \, \frac {\partial (\rho
v_i)}{\partial x_i}
\, = \, 0 \qquad .
\end{equation}

\item Momentum conservation.
\begin{equation}
\label{e:momentum}
\rho \frac{\partial v_i}{\partial t} + \rho v_j
\frac{\partial v_i}{\partial x_j}
\equiv \frac {\partial (\rho v_i)}{\partial t} +
\frac{\partial}{\partial x_j} \left( \rho v_i v_j \right)
=
- \rho \frac {\partial\phiG}{\partial x_i}
- \frac {\partial p}{\partial x_i}
+ \frac{\partial}{\partial x_j}
\left( \frac {B_iB_j}{4 \pi} - \delta_{ij} \frac {B^2}{8 \pi} \right) +
\frac {\partial \tv_{ij} }{\partial x_j} \qquad .
\end{equation}

\item The induction equation for magnetic flux conservation.
\begin{equation}
\label{e:B_induced}
\frac {\partial B_i}{\partial t} + \epsilon_{ijk} \epsilon_{klm} 
\frac {\partial}{\partial x_j}
\left( B_l v_m \right) = \eta \nabla^2 B_i \qquad ,
\end{equation}
which is derived from Maxwell's equations,
\begin{equation}
\label{e:Maxwell}
J_i = \frac {c}{4\pi} \epsilon_{ijk} \frac{\partial B_k}{\partial x_j}
\quad , \quad
-\frac {1}{c} \frac {\partial B_i}{\partial t} =
\epsilon_{ijk} \frac {\partial E_k}{\partial x_j}
\end{equation}
and a scalar conductivity law, 
\begin{equation}
J_i = \sigma \left( E_i + \frac{1}{c} \epsilon_{ijk} v_jB_k \right) 
 \qquad ,
\end{equation}
with $\sigma$ the conductivity and $\eta = c^2/4\pi \sigma$ the resistivity.

\item  Internal energy.
\begin{equation}
\label{e:E_int}
\frac {\partial u}{\partial t} +
\frac {\partial}{\partial x_i} (u v_i)  =
- p v_{i,i} - F_{i,i} - Q_{i,i}
+ \frac{J^2}{\sigma} + \tv_{ij}v_{i,j} \qquad .
\end{equation}

\item Electromagnetic energy.

We define the magnetic energy density, $\uB$, the magnetic stress tensor,
$\tB_{ij}$, and the fluid shear tensor, $s_{ij}$, by:
\begin{eqnarray}
\label{e:ub}
\uB &=& \frac {B^2}{8 \pi} \qquad , \\
\label{e:tBij}
\tB_{ij} &=& \frac {B_i B_j}{4 \pi} - \delta_{ij} \frac {B^2}{8 \pi} \qquad , \\
\label{e:s_ij}
s_{ij} &=& \frac {1}{2} \left( v_{i,j} + v_{j,i} - \frac {2}{3} \delta_{ij}
v_{k,k} \right) \qquad .
\end{eqnarray}
Taking the scalar product of the induction equation, (\ref{e:B_induced}),
with $B_i$ gives
\begin{equation}
\label{e:Emag}
\frac{\partial \uB}{\partial t} +
\frac{\partial}{\partial x_j} \left( \uB v_j\right) +
\frac{1}{3} \uB v_{k,k} 
=  \tB_{ij} s_{ij} + \frac{\eta}{4 \pi} B_i \nabla^2 B_i \qquad ,
\end{equation}
This equation describes the volume rate of change of magnetic energy due to
advection, expansion or compression, shearing and diffusion of field lines.
We write this equation in this particular form so as to emphasize the shear
term, which is the source term for magnetic field amplification in a fluid
with shearing motions, such as an accretion disk.
Equation (\ref{e:Emag}) also shows the well-known result that in the absence of shear or in the case of an isotropic
magnetic field the magnetic energy density  evolves similarly to a gas with adiabatic index
$\gamma = 4/3$ (i.e. pressure = 1/3 $\times$ energy density).
The diffusion term in (\ref{e:Emag}) satisfies
\begin{equation}
\frac {\eta}{4 \pi} B_i \nabla^2 B_i = - \frac{J^2}{\sigma}
- \frac {\partial^2  (\eta \tB_{ij})}{\partial x_i \partial x_j} 
\qquad ,
\label{e:B_diffusion}
\end{equation}
which describes the effects of Joule heating and the diffusion of field lines.
Equation~(\ref{e:Emag}) can then be expressed as
\begin{equation}
\frac {\partial \uB }{\partial t} +
\frac {\partial}{\partial x_i} \left[ \uB v_i
+ \eta \frac{\partial \tB_{ij} }{\partial x_j}
\right]
= \tB_{ij} s_{ij} - \frac{1}{3} \uB v_{k,k}
- \frac{J^2}{\sigma}
\qquad .
\label{e:Emag2}
\end{equation}
Note that the Poynting flux is
\begin{equation}
\label{e:Si}
S_i = \frac{c}{4\pi} \epsilon_{ijk} E_j B_k 
= \uB  v_i - \tB_{ij}  v_j + \eta \frac{\partial \tB_{ij}}{\partial x_j}
\qquad ,
\end{equation}
Equation~(\ref{e:Emag2}) can then be expressed in the more familiar form
\begin{equation}
\label{e:EM_cons}
\frac{\partial \uB}{\partial t} + S_{i,i}
= -J_iE_i = - \epsilon_{ijk} \frac{v_i}{c} J_j B_k - \frac{J^2}{\sigma}
\end{equation}
describing conservation of electromagnetic energy.

All of the above forms of the equations describing the evolution of the magnetic energy
in a fluid are useful.
Nevertheless, it is worth emphasizing that  when an induced current is involved
in $J_iE_i$ (as it is here through the expression for the current), the Poynting
flux does not correctly represent the flux of electromagnetic energy.
That is, the Poynting flux is only a good representation of the flux of electromagnetic
energy {\em in vacuo}.
This point that is not widely appreciated although
it has been known for some time in the plasma astrophysics of emission processes (see
for example, \citet{melrose91}).
This point assumes some importance when we consider the disk energy budget
in \S~\ref{s:energy_budget}.

\item Total energy. \\
\nobreak
We define the specific enthalpy of gas plus radiation by
\begin{equation}
h = \frac{u+p}{\rho}
\end{equation}
and we also define the total energy and corresponding energy flux by
\begin{eqnarray}
u_{\rm tot} &=&  \frac{1}{2} \rho v^2 + \rho \phiG + u + \uB
\label{e:e1}\\
\FE_i &=& \left( \frac{1}{2} \rho v^2 + \rho \phiG + \rho h \right) v_i
 + F_i + Q_i +  S_i  - \tv_{ij} v_j  \qquad .
 \label{e:fe1}
\end{eqnarray}
 The conservation equation for the total energy is obtained by taking the scalar
product of the momentum equation, (\ref{e:momentum}), with $v_i$ and then utilizing
both the internal energy equation, (\ref{e:E_int}), and the electromagnetic energy
equation~(\ref{e:EM_cons}) to obtain
 \begin{equation}
\label{e:etot}
\frac {\partial u_{\rm tot}}{\partial t} + \frac {\partial F^E_i}{\partial x_i}
= 0 \qquad .
\end{equation}
Equation~(\ref{e:fe1}) is the more usual form for the energy flux that explicitly
incorporates the Poynting flux.
Nevertheless, a more useful form for our purposes is derived by substituting the
expression~(\ref{e:Si}) for the Poynting flux, giving
\begin{equation}
\label{e:fe2}
\FE_i = \left( \frac{1}{2} \rho v^2 + \rho \phiG + \rho h  + \uB \right) v_i
-\tB_{ij}v_j + F_i + Q_i - \tv_{ij} v_j  + \eta \frac {\partial \tB_{ij}}{\partial x_j}
\qquad .
\end{equation}
In this expression, the Poynting flux is mainly replaced by an advection term $\uB v_i$ and a term $\tB_{ij} v_j$ representing the rate of work by magnetic stresses on the flow.

\end{enumerate}

\section{Statistical averaging -- the mean-field equations}

\label{s:mean_field}

\subsection{Definitions of mean variables}

In the following, we make extensive use of a statistical averaging approach 
\citep[e.g.][]{favre69,bradshaw76,krause80,bicknell84,bicknell86} in which the 
MHD equations  are ensemble-averaged.
When the flow is steady in the mean, we can think of the averaging process as
involving an average over a time-scale large compared to the time scales of
instabilities and turbulent fluctuations.
Statistical averaging is not commonly used in astrophysics ( at least not explicitly).
However, it is commonly used in  the theory of the physics of turbulent fluids
and in application areas such as geophysical fluid dynamics.
The approach is capable of providing valuable insights.
For example, in the case of accretion disks we gain valuable insights
into the energy flow within the system.
 
To describe the mean flow dynamics, we adopt the prescription whereby a quantity,
$X$, is expressed in terms of mean and fluctuating components and
$\langle X \rangle$ denotes its ensemble average value.
We distinguish intensive quantities, such as velocity, from extensive quantities,
such as density, pressure and magnetic flux density, by using mass-averaged values,
$\tilde X$, and mean values, $\bar X$, respectively.
This has the effect of conserving the mass associated with the mean flow, of
conserving the mean magnetic flux and of restricting the number of terms that
are produced by the averaging procedure.
Thus, the statistical averages of the density, velocity, pressure and magnetic
field are represented by:
\begin{equation}
\label{e:fluct_var}
\begin{array}{ c c l c c c l}
\rho & = & \bar \rho + \rho^\prime  & \qquad &  \langle \rho^\prime \rangle & =  & 0 \\
v_i & = & \tilde v_i + v_i' &  \qquad & \langle \rho v_i^\prime \rangle &  =  & 0 \\
p & = & \bar p + p^\prime & \qquad & \langle p^\prime \rangle & =  & 0 \\
B_i & = & \bar B_i + B_i^\prime  & \qquad & \langle B_i^\prime \rangle & = & 0 \\
\end{array}
\end{equation}
and the mean fluid shear is 
\begin{equation}
\label{e:sij_mean}
\tilde s_{ij} = \frac{1}{2} \left( \tilde v_{i,j} + \tilde v_{j,i}
- \frac{2}{3} \delta_{ij} \tilde v_{k,k} \right) \qquad .
\end{equation}

In this paper, we restrict ourselves to the simplest case in which the magnetic
field is dominated by its fluctuating component, so that $\langle B \rangle \equiv
\bar B \ll  \langle B^{\prime 2} \rangle^{1/2}$.
The more complex case in which a systematic component of the magnetic field
is present clearly has important implications for disk accretion solutions,
particularly when the field strength is sufficiently high to quench the MRI and
transport angular momentum via large-scale magnetic torques such as those invoked in
models for MHD-driven outflows and Poynting flux jets
\citep*[e.g.][]{blandpayne82,lovelace91,li92,wardkon93,ustyugova00}.
However, it is unclear whether large-scale, organized mean fields can be generated
from a highly chaotic underlying flow and whether they can explain collimated
outflows associated with accretion disks
\cite[see e.g. the discussion in][]{heinzbeg00}.
The more general case of $\bar B \neq 0$ is thus defered for future work.
All the mean-field equations derived here can, however, be straightforwardly
generalized to include terms involving $\bar B$ that are directly analogous to
their turbulent counterparts.

The molecular viscous stress tensor can also be written as
$\tv_{ij} = \bar \tv_{ij} + {\tv}_{ij}^\prime$, with
$\langle {\tv}_{ij}^\prime \rangle = 0$, since it can be related to a coefficient of
kinematic shear viscosity, $\nu$, by $\tv_{ij}=2\nu \rho s_{ij}$ (ignoring bulk
viscosity), where $s_{ij}$ is the shear tensor defined in (\ref{e:s_ij}).
In practice, the mean viscous stress, $\langle \tv_{ij} \rangle = \bar \tv_{ij}$,
usually has a negligible effect on momentum transport (particularly in a high
Reynolds number turbulent flow) and is thus usually ignored.
However, the fluctuating part of the viscous stress plays an important role
in the dissipation of turbulent energy on the smallest scales at the end of a
turbulent cascade.
Mathematically, this is described by the appearance of the correlation term
$\langle \tv_{ij} v_{i,j}^\prime \rangle$ in the mean-field internal energy equation,
(\ref{e:internal_mean}), derived below.
This term represents the mean rate of turbulent viscous heating and is dominated by the
high wavenumber components of the turbulent velocity fluctuations; that is, by the
dissipative region (in wavenumber space) of the turbulent cascade.
Thus, although we henceforth assume $\langle \tv_{ij} \rangle =0$, we retain
$\tv_{ij}$ throughout the averaging procedure in order to consistently
include stochastic energy dissipation by turbulent viscous stresses.

\subsection{Mass conservation}

Statistical averaging of the continuity equation, (\ref{e:mass}), gives
\begin{equation}
\frac{\partial \bar \rho}{\partial t} + 
\frac{\partial ( \bar \rho \tilde v_j )}{\partial x_j}= 0 \qquad ,
\label{e:mass_mean}
\end{equation}
so that the mass defined by the mean density is conserved.

\subsection{Momentum conservation}

In the case of zero nett magnetic flux, the magnetic stresses are
\begin{equation}
\tB_{ij} =  \frac {B_i' B_j'}{4 \pi} - \delta_{ij} \frac{B'^2}{8 \pi} \qquad .
\label{e:tmij}
\end{equation}
with ensemble mean,
\begin{equation}
 \langle \tB_{ij} \rangle  =
\frac{\langle B_i^\prime B_j^\prime  \rangle}{4 \pi}
- \delta_{ij} \frac {\langle {B^\prime}^2 \rangle}{8 \pi} 
 \end{equation}
We also define the Reynolds stresses as
\begin{equation}
\tR_{ij} = - \langle \rho v_i' v_j' \rangle
\label{e:tRij}
\end{equation}

Statistical averaging of the momentum equation, (\ref{e:momentum}),
yields
\begin{equation}
\frac{\partial ( \bar \rho \tilde v_i )}{\partial t} +
\frac{\partial \left( \bar \rho \tilde v_i \tilde v_j \right)}{\partial x_j}
=
- \bar \rho \frac{\partial \phiG}{\partial x_i}
- \frac{\partial \bar p}{\partial x_i}
+ \frac{\partial}{\partial x_j}
\left( \tR_{ij} +  \langle \tB_{ij} \rangle \right) \qquad .
\label{e:motion_mean}
\end{equation}
This equation shows that momentum transport in a fluctuating MHD fluid involves
additional, statistically-averaged quantities that arise from the turbulent magnetic
stresses and the Reynolds stresses.
Note that the Reynolds stress is conventionally defined as a mean quantity, whereas,
for convenience,
we have defined $\tB_{ij}$ as an unaveraged quantity in order to simplify
triple correlation terms that appear in the energy equations derived in
\S~\ref{s:Energy} below.
Note also that these definitions of the turbulent stresses are opposite in sign
to those used, for example, by \citet{balbhaw98}.
Our usage is consistent with the conventional description of this term as a stress.

Often in astrophysical contexts, the terms $\tR_{ij}$ and $\tB_{ij}$ are referred to as
viscous and resistive terms since they represent additional forms of momentum and
electromagnetic transport.
In this paper, we consistently refer to these terms as turbulent stresses and reserve
the term viscosity and resistivity for real molecular effects.

\subsection{The induction equation}

Although we assume that the mean magnetic flux density is zero, for completeness,
and to make one physical point, we present here the mean-field induction
equation obtained by statistically averaging (\ref{e:B_induced}):
\begin{equation}
\frac {\partial \bar B_i}{\partial t} +
\epsilon_{ijk} \epsilon_{klm} \frac {\partial}{\partial x_j}
\left( \bar B_l \tilde v_m 
+ \langle B^\prime_l v^\prime_m \rangle \right) = \eta \nabla^2 \bar B_i
\end{equation}
This equation forms part of the foundation of classical mean-field electrodynamics
\citep[e.g.][]{krause80}, which is usually developed under the assumption of
an incompressible flow.
The contribution from the term involving $\bar B_l \tilde v_m$ results from
compressibility, while the term involving $\langle B^\prime_l v^\prime_m \rangle$
gives rise to dynamo amplification resulting from mean helicity.
As exceptionally clarified by \citet[][ see Sec.~VI-B]{balbhaw98}, when a
weak seed field $\bf B$ is present in a differentially rotating fluid, the
fluctuating fields $\bf v^\prime$ and $\bf B^\prime$ are self-consistently
generated by the MRI and thus, cannot be prescribed independently of $\bf B$,
which is the underlying assumption of mean-field dynamo theory.
We do not, therefore, make any assumption in our formalism concerning
mean-field dynamo amplication.

\subsection{Energy}
\label{s:Energy}

There are a number of different energies to account for in a turbulent MHD 
fluid: the turbulent magnetic and kinetic energies, the internal energy and the
total (turbulent plus internal plus mechanical) energy.
Here, we derive the conservation equations describing the evolution of
these quantities.

\subsubsection{Turbulent magnetic energy}
\label{s:TME}

The mean-field equation for the turbulent magnetic energy density,
$\langle \uB \rangle  = \langle B^{\prime 2} \rangle/8\pi$, is derived
by statistically averaging the electromagnetic energy equation, (\ref{e:Emag2}),
yielding
\begin{equation}
\frac {\partial}{\partial t} \langle \uB \rangle +
\frac {\partial}{\partial x_i} \left[  \langle \uB \rangle \tilde v_i
+ \langle \uB v^\prime_i \rangle \right]
= \langle \tB_{ij} \rangle \tilde s_{ij}
+ \langle \tB_{ij} s^\prime_{ij} \rangle
- \frac {1}{3} \langle \uB \rangle \tilde v_{k,k}
- \frac {1}{3} \langle \uB  v^\prime_{k,k} \rangle
- \eta \frac{\partial^2 \langle \tB_{ij} \rangle}{\partial x_i \partial x_j}
- \frac{\langle J^2 \rangle}{\sigma}
\qquad ,
\label{e:TME}
\end{equation}
where the turbulent magnetic stress tensor, $\tB_{ij}$ is defined by (\ref{e:tmij}).
The first term on the right hand side of (\ref{e:TME}) is a source
term for the production of turbulent magnetic energy as a result of interaction
between the turbulent magnetic stresses and the mean fluid shear.
The second term represents the interaction between the turbulent magnetic stresses
and the fluctuating components of the mean shear.
The third and fourth terms describe changes in the turbulent magnetic energy due
to compression and expansion in the mean and fluctuating components of the flow,
respectively.
The last two terms on the right hand side of (\ref{e:TME}) describe the total
work done by resistive forces minus the rate at which turbulent magnetic energy
is converted into heat via Joule losses.

\subsubsection{Turbulent kinetic energy}
\label{s:TKE}

A similar mean-field equation for the turbulent kinetic energy density,
$\langle \uK \rangle = \langle \frac{1}{2} \rho v^{\prime 2} \rangle$, can be
derived by taking the scalar product of $v_i$ with the unaveraged momentum equation,
(\ref{e:momentum}), statistically averaging this equation  and then subtracting from
it the scalar product of $\tilde v_i$ with the statistically averaged momentum equation,
(\ref{e:motion_mean}).
In symbolic terms, this is $\langle {\bf v} \cdot \hbox {momentum eqn.}
\rangle - \tilde {\bf v} \cdot \langle \hbox{momentum eqn.}\rangle$,
which is equivalent to $\langle {\bf v^\prime}\cdot \hbox{momentum eqn.} \rangle$
and yields
\begin{eqnarray}
\frac{\partial}{\partial t} \langle \uK \rangle &+&
\frac{\partial}{\partial x_i} \left[ \langle \uK \rangle
\tilde v_i + \langle \uK v_i^\prime \rangle +
\langle p v^\prime_i \rangle - \langle \tB_{ij} v^\prime_j \rangle -
\langle \tv_{ij} v^\prime_j \rangle \right] \nonumber \\
&=&   \tR_{ij} \tilde s_{ij}
-  \frac {2}{3} \langle \uK \rangle \tilde v_{k,k}
- \langle \tB_{ij} s^\prime_{ij} \rangle
+ \langle \frac{1}{3}\uB v_{k,k}^\prime \rangle
+ \langle p v^\prime _{i,i} \rangle
- \langle \tv_{ij} v^\prime_{i,j} \rangle \qquad .
\label{e:TKE}
\end{eqnarray}
Note the similarity to the turbulent magnetic energy equation, (\ref{e:TME}), and also to the
internal energy equation for an adiabatic $\gamma=5/3$ gas when the
triple correlations are neglected.
In particular, there is an analogous source term, $\tR_{ij}\tilde s_{ij}$,
describing the rate at which shear in the mean flow does work on the
Reynolds stresses.
There is also a sink term, $\langle \tv_{ij} v^\prime_{i,j} \rangle$,
describing viscous dissipation of turbulent kinetic energy into heat.
Note also that the turbulent kinetic energy equation is coupled directly
to the turbulent magnetic energy equation
via the magnetic term on the right hand side of  (\ref{e:TKE}), which
also appears explicitly on the right hand side of (\ref{e:TME}).

\subsubsection{Total turbulent energy}
The magnetic coupling term in (\ref{e:TKE}) implies that the turbulent kinetic
and magnetic equations
can be combined into a single turbulent energy equation:
\begin{eqnarray}
\frac{\partial}{\partial t} \langle \uK + \uB \rangle
& + &
\frac{\partial}{\partial x_i} \left[ \,
\langle \uK + \uB \rangle \tilde v_i
+ \langle \left( \uK + \uB \right) v_i^\prime \rangle
+ \langle \left( \delta_{ij}p -  \tB_{ij} - \tv_{ij} \right) v^\prime_j \rangle
+  \eta \frac{\partial \langle \tB_{ij} \rangle}{\partial x_j}
\, \right] \hspace{1.0truecm} \nonumber \\
& =&   \langle t_{ij} \rangle \tilde s_{ij}
- \langle \frac{2}{3}\uK + \frac{1}{3}\uB \rangle \tilde v_{k,k}
+ \langle p v^\prime _{i,i} \rangle
- \langle \tv_{ij} v^\prime_{i,j} \rangle 
- \frac{\langle J^2 \rangle}{\sigma}
\qquad ,
\label{e:TE}
\end{eqnarray}
where 
\begin{equation}
\langle t_{ij} \rangle = \tR_{ij} + \langle \tB_{ij} \rangle
\label{e:tij}
\end{equation}
is the combined Reynolds and magnetic stress tensor.

The last two terms on the right hand side of (\ref{e:TE}) represent the dissipation
of turbulent kinetic and magnetic energy whereas the first term on the right represents
the {\em production} of turbulent energy through the action of the mean shear on the
total stresses.
The usual approach in many physical applications is to equate the
heating rate to the production term.
However, the presence of transport terms, as well as terms describing the work
done by fluid compression and/or expansion, implies that the steady-state production
and dissipation of turbulent energy are non-local and therefore, the rates cannot in
general be equated.
Thus, it is not generally correct to simply replace the viscous dissipation term
in the internal energy equation with the production term for turbulent energy.
The rate of turbulent viscous dissipation in the mean-field internal energy equation
must be related to the rate of production of turbulent energy self-consistently
using (\ref{e:TE}), as is demonstrated in the following section.

\subsection{Internal energy}
\label{s:IntE}

Statistically averaging the internal energy equation, (\ref{e:E_int}), yields
\begin{equation}
\frac{\partial \bar u}{\partial t} +
\frac{\partial}{\partial x_i} \left(
\bar u \tilde v_i 
+ \langle u  v_i^\prime \rangle \right) =
-\bar p \tilde v_{i,i} - \langle p v_{i,i}^\prime \rangle
- \langle F_{i,i} \rangle - \langle Q_{i,i} \rangle 
+ \langle \frac {J^2}{\sigma} \rangle
+ \langle \tv_{ij} v_{i,j}^\prime \rangle \qquad .
\label{e:internal_mean}
\end{equation}
The terms on the left hand side describe the total rate of change in the gas plus
radiation internal energy density as a result of intrinsic temporal variations
and advective transport by the mean plus turbulent flow.
The terms on the right hand side of (\ref{e:internal_mean}) describe the work done
by compression or expansion in the flow against the gas and radiation pressure,
radiative losses, energy exchange from an external heat source or sink, mean-field
Ohmic heating and turbulent viscous heating.
The source terms determine the rate at which energy is converted into random
particle energy (some of which is then converted into radiation) and also
into bulk kinetic energy.
Magnetic energy in particular can be converted directly via Joule heating
(usually identified with field line reconnection) as well as via work done by the
turbulent flow against the turbulent viscous stresses

\subsection{Turbulent plus internal energy}

The viscous and Joule dissipation terms appear as sink terms in the
turbulent  energy equation (\ref{e:TE}) and as source terms in the internal energy
equation (\ref{e:internal_mean}).
Hence, combining these equations eliminates these terms with the following result
for the total internal energy of a turbulent MHD fluid:
\begin{eqnarray}
\frac{\partial}{\partial t}
\left( \bar u + \langle \uK + \uB \rangle \right)
+ \frac{\partial}{\partial x_i} \left[
\left(\bar \rho \tilde h + \langle \uK + \uB \rangle \right) \tilde v_i 
+ \langle \rho h^\prime v^\prime_i \rangle + \langle \left( \uK + \uB  \right) v_i^\prime \rangle
- \langle \left( \tB_{ij} + \tv_{ij} \right) v^\prime_j \rangle
+ \eta \frac{\partial \langle \tB_{ij} \rangle}{\partial x_j}
\right]  \nonumber \\
=
 \langle t_{ij} \rangle \tilde s_{ij}
- \langle \frac{2}{3}\uK + \frac{1}{3}\uB \rangle \tilde v_{k,k}
- \bar p  \tilde v_{i,i}
- F_{i,i} - Q_{i,i}  
\qquad .
\label{e:total_int_mean}
\end{eqnarray}

\subsection{Total energy}
\label{s:TotE}

Statistically averaging the total energy equation, (\ref{e:etot}), gives
\begin{equation}
\frac {\partial \bar u_{\rm tot}}{\partial t}
+ \frac {\partial \langle  \FE_i \rangle }{\partial x_i} = 0
 \qquad ,
\label{e:energy_mean}
\end{equation}
where the average total energy is:
\begin{equation}
\bar u_{\rm tot} = \frac {1}{2} \bar \rho \tilde v^2 + \bar \rho \phiG
+ \langle u \rangle + \langle \uK \rangle + \langle  \uB \rangle
\end{equation}
and involves both the components that one expects from the mean flow plus the average
turbulent kinetic energy, $\langle \uK \rangle $.
We use the second form of the energy flux, equation~(\ref{e:fe2}), to form the mean
energy flux:
\begin{eqnarray}
\langle \FE_i\rangle &=& \left[ \frac {1}{2} \bar \rho \tilde v^2 +  \bar \rho \phiG 
+ \bar \rho \tilde h   + \langle \uK \rangle + \langle \uB \rangle   \right] \tilde v_i
\nonumber \\
&& +  \langle  \rho h^\prime v^\prime_i \rangle + \langle  \uK v_i^\prime \rangle 
+  \langle  \uB v^\prime_i  \rangle
- \langle t_{ij}  \rangle \tilde v_j - \langle t_{ij} v^\prime_j \rangle \nonumber \\
&& + \langle F_i \rangle + \langle Q_i \rangle  
- \langle \tv_{ij} v^\prime_j \rangle
+ \frac {\partial \eta \langle \tB_{ij} \rangle}{\partial x_j}
\label{e:Fe2}
\end{eqnarray}

The numerous terms in this expression for the mean energy flux, $\langle \FE_i \rangle$, are all easily interpreted.
The first group of five terms represents the energy flux advected by the mean flow and consists of bulk kinetic, gravitational, enthalpy, kinetic and magnetic terms.
The next group of three terms
($ \langle  \rho h^\prime v^\prime_i \rangle + \langle  \uK v_i^\prime \rangle +  \langle  \uB v^\prime_i  \rangle$) represent the turbulent fluxes of enthalpy, turbulent kinetic energy and turbulent magnetic energy.
The next two terms ($- \langle t_{ij}  \rangle \tilde v_j - \langle t_{ij} v^\prime_j \rangle$) represent the nett work done by the  {\em  total} Reynolds plus magnetic stresses on the mean and turbulent flow.
The next two terms ($ \langle F_i \rangle + \langle Q_i \rangle$)  represent the
radiative and heat fluxes respectively.

The last two terms ($- \langle \tv_{ij} v^\prime_j \rangle + \langle \eta \partial \tB_{ij} \rangle_{,j} $) represent the work done by the viscous and resistive stresses. In most regions of a high  Reynolds number flow, they 
are negligible.
Nevertheless, in some regions of high spatial gradients, such as a shock or a reconnection region, they could be
comparable to the other terms and their inclusion in equation~(\ref{e:Fe2}) is logical
\citep[see e.g.][ for further details on stochastic reconnection]{lazvish99}.
However, when the energy equation is integrated over a volume whose bounding surface is well outside regions of dissipation, the contribution of these two terms to the resulting surface integral of the energy flux is negligible.  Therefore, in many circumstances, we can safely ignore these terms in considering the integral form of the energy equation.
This point is elaborated in \S~\ref{s:energy_budget} in the context of accretion disks.

The total energy equation  describes the nett transfer of energy from one component to another and incorporates the work done by the total turbulent stresses through the  term $\langle t_{ij} \rangle \tilde v_j$
as well as the change in binding energy represented by the terms $\case{1}{2} \bar \rho \tilde v^2 + \bar \rho \phiG$.
The volume dissipative terms ($\langle \tv_{ij} s_{ij} \rangle$ and $\langle J^2/\sigma \rangle$) disappear because of their equal and opposite contributions in the equations describing production of turbulent kinetic and  magnetic energy on one hand and dissipation of that energy into heat on the other. Other terms in the energy equation represent non-local effects such as advection and diffusive transport.

Note that the advection terms contain contributions from both turbulent kinetic and magnetic energy, not just enthalpy. Hence, in applications to advective accretion disk models, for example, the magnetic and turbulent kinetic energy should be taken into account especially when the magnetic field is near equipartition or when the turbulent velocities are near transonic.

\subsection{Poynting flux}

For future reference (in \S~\ref{s:energy_budget}), we also note the mean Poynting flux:
\begin{equation}
\langle  S_i \rangle = \langle \uB \rangle \tilde v_i + \langle \uB v^\prime_i \rangle  
- \langle \tB_{ij} \rangle  \tilde v_j - \langle \tB_{ij}  v_j ^\prime \rangle 
\label{e:poynting}
\end{equation}

\subsection{Mechanical energy of the mean flow}

For completeness, we note that an equation for the mechanical energy can be derived either by
taking the scalar product of the momentum equation, (\ref{e:motion_mean}),
with the mean velocity $\tilde v_i$, or by subtracting the
total internal energy equation, (\ref{e:total_int_mean}), from 
(\ref{e:energy_mean}), giving
\begin{equation}
\frac {\partial}{\partial t} \left[ \frac{1}{2} \bar \rho \tilde v^2 + \bar \rho \phiG \right] +
\frac{\partial}{\partial x_i}
\left[ \left(
\frac{1}{2} \bar \rho \tilde v^2
+ \bar \rho \phiG
\right) \tilde v_i
\right]
+ \tilde v_i \left( \frac{\partial \bar p}{\partial x_i}
- \frac{\partial \langle t_{ij} \rangle }{\partial x_j} \right) = 0 \qquad .
\label{e:energy_mech}
\end{equation}

\section{Accretion Disk Dynamics}
\label{s:disc_eqns}
 
We now apply the generalized mean-field equations derived in the preceding section
to an accretion disk around a black hole. In this section we consider the implications of the radial, azimuthal and vertical momentum equations. We then apply these results to the energy budget in acccretions disk in \S~\ref{s:energy_budget}.

In the following, we present the full statistically-averaged equations in
a cylindrical $(r,\phi,z)$ coordinate system for a fluid which is time-independent
and axisymmetric in the mean
$(\langle \partial / \partial t \rangle = \langle \partial /\partial \phi \rangle =0)$.
We use Newtonian physics throughout, with the gravitational
potential $\phiG = -GM(r^2 + z^2)^{-1/2}$.
We integrate the mean-field conservation equations vertically over an
arbitrary disk scaleheight, $h = h(r)$.
Quantities calculated at the disk surface ($z= \pm h$) are denoted by a $\pm$
superscript, $X^{\pm}$, and we assume reflection symmetry about the disk midplane,
so that $|X^+| = |X^-|$.
Midplane values of variables are denoted by $X_0$.

Since we include the effects of a disk wind, we do not identify $h$ as a
hydrostatic scaleheight, but  as a photospheric height,
delineating between the disk proper and the transition region leading
to a corona, by analogy with the solar atmosphere.
We assume $h$ is much less than the radius, so that quantities of order
$h/r$ and ${\rm d}h/{\rm d}r$ are neglected.
The following comments are useful in the context of the vertical integration
process.
Let us represent a generic conservation law for mass, momentum, energy etc. by
\begin{equation}
\frac {1}{r} \frac {\partial (r A_r)}{\partial r} + \frac {\partial
A_z}{\partial z} = S \qquad .
\end{equation}
Multiplying by $2 \pi r$ and integrating with respect to the
vertical coordinate, $z$, from $-h$ to
$h$ gives:
\begin{equation}
\int_{-h}^{+h} \frac {\partial (2 \pi r A_r)}{\partial r} \> \dz
+ 2 \pi r (A_z^+ - A_z^-) = \int_{-h}^{+h}  2 \pi r \, S \> \dz
\end{equation}
Taking the derivative with respect to $r$ outside of the integral gives:
\begin{equation}
\dr \int_{-h}^{+h} 2 \pi r A_r  \> \dz - 2 \pi r (A_r^+ + A_r^-)
\frac{{\rm d}h}{{\rm d}r} + 2 \pi r (A_z^+ - A_z^-) = \int_{-h}^{+h} 2 \pi r \, S \> \dz
\qquad .
\label{e:generic}
\end{equation}
Since reflection symmetry about the midplane holds, 
\begin{equation}
|A_r^+| = |A_r^-| \qquad\hbox{and}\qquad |A_z^+| = |A_z^-| \qquad .
\end{equation}
Generally, there are two surface terms resulting from the integration
over disk height, the second
and third terms on the left of equation~(\ref{e:generic}). Unless
$|A_r| >> |A_z|$, the first surface
term, which is proportional to ${\rm d}h/{\rm d}r = {\cal O} (h/r)$ is negligible
compared to the second, so that
the result of the integration over $z$ is:
\begin{equation}
\dr \int_{-h}^{+h} 2 \pi r A_r  \> \dz
+ 2 \pi r (A_z^+ - A_z^-) \simeq \int_{-h}^{+h} 2 \pi r \, S \> \dz
\label{e:vertical}
\end{equation}

We believe the geometrically-thin disk approximation \citep{lbpring74}
provides the most physically plausible disk solutions, given that
geometrically-thick, advection-dominated disks are founded upon assumptions
(namely preferential ion heating, negligible ion--electron coupling,
and negligible electron heating overall) that are highly idealized,
particularly in the presence of MHD turbulence
(e.g. \citealt{bkl97,bkl00};  see also \citealt{merlfab02}).
Furthermore, \citet{begelman02} has recently demonstrated that
radiation-pressure-dominated solutions need not be concomitant with
geometrically-thick disks either.

Since the MRI is a weak-field instability, it drives subsonic turbulence,
with $ \langle \rho v^{\prime 2} \rangle
\ltapprox \langle \rho c_{\rm s}^2 \rangle \ll \bar \rho \OmegaK$, where
$c_{\rm s} = (kT/\mu m_{\rm p})^{1/2}$
is the local sound speed and $\OmegaK = (GM/r^3)^{1/2}$ is the Keplerian angular
velocity.
An important distinction between our approach and that adopted in numerical
simulation models (see \citealt{balbhaw98}) is that we do not identify the
mean and fluctuating fluid velocity components with the Keplerian and 
non-Keplerian components, respectively.
Instead, we adopt the more formal statistical
averaging approach (as outlined in Sec.~\ref{s:mean_field}) of decomposing
all velocity components into mean and fluctuating parts.
Thus, we explicitly distinguish between mean radial, azimuthal and vertical motions
and their fluctuating counterparts and only the fluctuating components are restricted
to subsonic speeds.
The only restriction we place on the mean fluid velocity components is that they
satisfy $\tilde v_\phi \gg \tilde v_r , \tilde v_z$ and that
$\tilde v_r$ and $\tilde v_\phi$ are independent of $z$, simplifying the
vertical integration.
Thus, we are able to self-consistently treat radial and vertical transport of mass,
momentum and energy by the mean fluid velocity fields.

\subsection{Mass transfer}
\label{s:mass}

Vertical integration of the mean-field continuity equation,
(\ref{e:mass_mean}),
\begin{equation}
\label{e:cont_mean}
\frac{1}{r}\frac{\partial (r\bar \rho \tilde v_r)}{\partial r} +
\frac{\partial (\bar \rho \tilde v_z)}{\partial z} = 0 \qquad ,
\end{equation}
gives
\begin{equation}
\dr \int_{-h}^{+h} 2 \pi r \bar \rho \tilde v_r \> \dz
\, + \, 4 \pi r \bar \rho^+ \tilde v_z^+  = 0  \qquad .
\label{e:mass_int}
\end{equation}
We now introduce the usual definitions for the surface mass density,
\begin{equation}
\Sigma(r) \equiv \int_{-h}^{+h} \bar  \rho \> \dz
\label{e:sigma_defn}
\end{equation}
and mass accretion rate,
\begin{equation}
\Mdota (r) = 2\pi r  \Sigma (- \tilde v_r)  \qquad .
\label{e:Mdota}
\end{equation}
We also introduce an analogous mass outflow rate,
\begin{equation}
\Mdotw (r) = \int_r^\infty 4\pi r \bar \rho^+ \tilde v_z ^+{\rm d}r
\, = \, \Mdotw (\ro) - \int_{\ro}^{r} 4\pi r \bar \rho^+ \tilde v_z ^+{\rm d}r
\label{e:Mdotw}
\end{equation}
associated with a mean vertical velocity $\tilde v_z^+$ at the disk 
surface, i.e. at the base of a wind.

Using the above definitions, the vertically integrated continuity equation,
(\ref{e:mass_int}), can be written as
\begin{equation}
\dr \Mdota (r) = 4\pi r \bar \rho^+ \tilde v_z ^+ = - \dr \Mdotw (r)
\label{e:dM}
\end{equation}
implying that
\begin{equation}
\Mdota (r) + \Mdotw (r) = \Mdota (\ro) + \Mdotw (\ro) = \hbox{constant} = \Mdot
\qquad ,
\label{e:Mdot}
\end{equation}
where $\Mdota (\ro) + \Mdotw (\ro)$ is the total mass flux at the innermost
stable orbit, $\ro$.
Equation (\ref{e:dM}) implies that under steady--state conditions,
the radial mass flux decreases towards small $r$ at the same rate as the
vertical mass flux increases in order to maintain a constant nett mass flux,
$\Mdot$, which is equivalent to the nett accretion rate at $r = \infty$.

\subsection{Radial momentum}
\label{s:rad_mom}
The radial component of the mean-field momentum equation, (\ref{e:motion_mean}),
is
\begin{equation}
\label{e:rad_mom0}
\frac {1}{r} \left[ \frac{\partial}{\partial r} \left(
r \bar \rho \tilde v_r^2 \right)  - \bar \rho \tilde v_\phi^2 \right]
+ \frac {\partial}{\partial z} \left( \bar \rho \tilde v_r \tilde v_z
\right)=
- \bar \rho \frac{\partial \phiG}{\partial r}
- \frac {\partial \bar p}{\partial r}
+ \frac{1}{r} \left[ \frac{\partial}{\partial r}
\left( r \langle t_{rr} \rangle \right) - \langle t_{\phi \phi} \rangle \right]
+ \frac {\partial \langle t_{rz} \rangle}{\partial z}
\end{equation}
Several of the terms in this equation are negligible compared to the
dominant term, $\bar \rho \tilde v_\phi^2 /r$.
To determine which terms can be neglected, we first use the mean-field continuity
equation, (\ref{e:cont_mean}), to simplify the $\partial /\partial r$ and
$\partial / \partial z$ terms on the left hand side.
Then, evaluating the gravitational term on the right hand side gives
\begin{equation}
\label{e:rad_mom}
\frac{\bar \rho \tilde v_\phi^2}{r}
- \frac{1}{2}\bar \rho \frac{\partial \tilde v_r^2}{\partial r}
=
\frac{GM\bar \rho}{r^2}
+ \frac {\partial \bar p}{\partial r}
- \frac{1}{r} \left[ \frac{\partial}{\partial r}
\left( r \langle t_{rr} \rangle \right) - \langle t_{\phi \phi} \rangle \right]
- \frac {\partial \langle t_{rz} \rangle}{\partial z}
\qquad .
\end{equation}
The radial gradient terms can be dropped since they are negligible
compared to $\bar \rho \tilde v_\phi^2/r$.
However, the vertical gradient in the turbulent stress, $\partial \langle t_{rz} \rangle / \partial z$,  could be
important in a geometrically thin disk, so this term is retained.
Integrating the remaining terms in (\ref{e:rad_mom}) over $z$ gives
\begin{equation}
\frac{\Sigma \tilde v_\phi^2}{r} \simeq \frac{GM\Sigma}{r^2}
-  2 \langle t_{rz} \rangle^+  \qquad ,
\end{equation}
which reduces to
\begin{equation}
\tilde v_\phi^2 \simeq \vK^2 - \frac{2r}{\Sigma}
\langle t_{rz} \rangle^+ \qquad .
\end{equation} 
Thus, Keplerian rotation prevails  in regions where the 
turbulent $rz$ stresses on the disk surface are $\ltapprox h_{\rm av}/r$ times smaller than
$\bar \rho \vK^2$, where $h_{\rm av}$ is the density scale height defined below (see equation~(\ref{e:h_av})).
Since the turbulent stresses saturate at quasi-thermal levels, they are
unlikely to modify the Keplerian profile.
Interestingly, models for flux emergence from the solar photosphere indicate
that the Parker instability can enhance the poloidal flux sufficently
to modify the background shear in the rotation velocity, thus giving rise
to an effective buoyant-shear instability \citep*[e.g.][]{cline03}.
In numerical simulations for accretion disks, however, this does not appear to
be the case, with Keplerian rotation profiles emerging in all models
\citep[e.g.][]{hawbalb02}.

\subsection{Vertical momentum}
\label{s:vertical_bal}

Vertical momentum balance in the disk is obtained from the
$z$-component of eqn.~(\ref{e:motion_mean}):
\begin{equation}
\frac{1}{r} \frac{\partial}{\partial r} \left( r \bar \rho \tilde v_r
\tilde v_z \right)
+ \frac {\partial}{\partial z} \left( \bar \rho \tilde v_z^2 \right)
=
-\frac {GM \bar \rho z}{r^3}
- \frac {\partial}{\partial z}
\left( \bar p   - \langle t_{zz} \rangle \right)
+ \frac{1}{r} \frac{\partial }{\partial r} \left( r \langle t_{rz} \rangle \right)
\qquad .
\label{e:vert_bal0}
\end{equation}
We define a density averaged disk height and a density averaged vertical velocity by:
\begin{equation}
h_{\rm av} = \frac{2}{\Sigma} \int_0^h \bar \rho z \> \dz \qquad , \qquad
\tilde v_{z,\rm av} = \frac{2}{\Sigma}\int_0^h \bar \rho \tilde v_z \> \dz
\qquad .
\label{e:h_av}
\end{equation}
Integration of equation~(\ref{e:vert_bal0}) over $z$ then yields:
\begin{equation}
\frac{1}{4\pi r}\frac{\partial}{\partial r}
\left( -\Mdota \tilde v_{z,\rm av} \right) + \bar \rho^+ \tilde v_z^{+2}
\simeq - \frac{GM\Sigma h_{\rm av}}{2r^3}
 \bar p_0 - \bar p^+ 
+ \langle t_{zz} \rangle^+ - \langle t_{zz} \rangle_0
\qquad ,
\label{e:vert_bal}
\end{equation}
where the $rz$ stress term has been dropped since its contribution is
$\sim h_{\rm av}/r$ that due to the vertical pressure gradients.

In order to determine the relative importance of the remaining terms,
note that the first term on the left hand side of (\ref{e:vert_bal}) can
be written as
\begin{equation}
\frac{1}{4\pi r}\frac{\rm d}{{\rm d} r}
\left( -\Mdota \tilde v_{z,\rm av} \right) = 
-\bar \rho ^+ \tilde v_z^+ \tilde v_{z,\rm av}
- \frac{1}{2}\Sigma (-\tilde v_r)\frac{{\rm d}\tilde v_{z,\rm av}}{{\rm d}r}
\end{equation}
where equations~(\ref{e:Mdota})--(\ref{e:dM}), involving the mass fluxes, have been used.
The first term on the right hand side of this equation is
$\tilde v_{z,\rm av} /\tilde v_z^+ \ll 1$ times smaller than the
$\bar \rho \tilde v_z^{+2}$ term in (\ref{e:vert_bal}) and is thus negligible.
The relative importance of the second term on the right hand side of the
above equation depends on the radial gradient of $\tilde v_z$; if we reasonably
suppose that $\partial \tilde v_z/\partial r \sim \tilde v_z /r$, then this
term is smaller than the $\bar \rho \tilde v_z^{+2}$ term in the vertical
balance equation, (\ref{e:vert_bal}), by a factor $\sim h/r$.
We thus neglect altogether the contribution from the first term on the left hand
side of (\ref{e:vert_bal}) to the overall vertical momentum transport in the disk.
Thus, vertical momentum balance in the disk implies a mean vertical
outflow from the disk surface given by
\begin{equation}
\tilde v_z^{+2} \simeq
 \frac{\bar p_0 - \bar {p}^+
+ \langle t_{zz} \rangle^+ - \langle t_{zz} \rangle_0}{\bar \rho^+} 
-\frac{GM\Sigma h_{\rm av}}{2\bar \rho^+ r^3}
\qquad .
\label{e:vz}
\end{equation}
 
The following well-known order of magnitude estimate for the hydrostatic equilibrium disk height is useful and we repeat it here for the sake of completeness. Equation~(\ref{e:vz}) with $\tilde v_z = 0$ and $\Sigma \sim 2 \rho_0 h_{\rm av}$ ($\rho_0 =$ central disk density),
gives
\begin{equation}
\frac {h_{\rm av}}{r} \sim \frac {c_0^{\rm tot}}{\vK}
\label{e:h}
\end{equation}
where $c_0 = \sqrt{p_0^{\rm tot}/\rho_0} $ is the generalized sound speed corresponding to the total 
central disk pressure $p_0^{\rm tot}= p_0 +  \langle \uB \rangle_0$. If the departures from equilibrium values of pressure, magnetic field etc. are by no more than a factor of order unity, then the velocity at the base of the wind, $\tilde v_z^+$ implied by (\ref{e:vz}) is of order $c_0$.

\subsection {Conditions for a disk wind}

In the context of the specific accretion disk application being considered here, a non-advective,
zero nett magnetic flux disk, with turbulence driven by the MRI instability, the
thermal pressure cannot drive an energetically-significant, large-scale wind,
not even for a coronal temerature $\sim 100\,$keV.
For example, by examining the Bernoulli equation (without magnetic field), it is straightforward to
show that the thermal temperature required to drive an outflow in the vicinity of a
black hole is $kT > \case{1}{2}(\gamma-1) GM \mu m_p /r = \case{1}{2}v_{\rm esc}^2
\simeq 10^5 (\rg/r) \> \rm keV$, where $\gamma$ is the adiabatic index,
$v_{\rm esc}$ is the escape speed, and $\rg = GM/c^2 \simeq 1.5 \times 10^{12}M_7\,$cm
is the gravitational radius of a black hole of mass $M=10^7M_7 M_\odot$.
If the disk luminosity is super-Eddington then a radiation pressure driven outflow is feasible  \citep{begelman02}. On the other hand  if the radiation field is sufficiently diffusive to smear out vertical gradients, then radiation pressure may be unimportant \citep{blaes02}. 

The most probable driving source for a large-scale wind is the magnetic field and one may derive approximate conditions for the production of a wind as follows.
Let $\vA$ be the Alfv\'{e}n speed, with mean value defined by
\begin{equation}
\vA = \langle \vA^2 \rangle^{1/2} = \left( \frac {\langle B^{\prime 2} \rangle }{4 \pi \bar \rho}\right)^{1/2} 
\end{equation}
and consider the vertical component of the energy flux at the disk surface without turbulent diffusion terms:
\begin{equation}
\langle \FE_z \rangle^+ \simeq \left(\frac {1}{2} \tilde v^{+2} + \phiG + \tilde h^+ 
+ \frac {\langle  B^{\prime 2}_r + B^{\prime 2}_\phi  \rangle^+}{4 \pi \bar \rho^+}   \right)
\bar \rho^+ \tilde v_z^+  
- \frac {\langle B_\phi^\prime B_z^\prime  \rangle^+ }{4 \pi} \,  \vK
\qquad .
\end{equation}
The azimuthal magnetic field component, $B_\phi^\prime$ should dominate so that we put
$\case  {\langle  B^{\prime 2}_r + B^{\prime 2}_\phi  \rangle^+}{4 \pi \bar \rho^+} \simeq \langle \vA^2 \rangle$. There are negative ($\case{1}{2}\tilde v^2+\phiG = -\case{1}{2} \case{GM}{r} $) and positive ($\tilde h^+$, $\langle \vA^2 \rangle^+ $) contributions to the first bracket of terms in the energy flux. The last term (the centrifugal term) makes a positive contribution to the energy flux if the magnetic stress,  
$(4 \pi)^{-1} \langle B_z^\prime  B_\phi^\prime  \rangle^+ < 0$.
In a steady state, the energy flux, integrated over the extent of the wind, is conserved and in order that the wind escape to infinity, the integrated energy flux should be positive. Neglecting, in the first instance,  the centrifugal term and disregarding the specific enthalpy, the energy flux is positive if
\begin{equation}
\vA  \gtapprox \frac {1}{\sqrt 2} \vK \qquad ,
\label{e:wind_condn1}
\end{equation}
that is, the Alfven speed in the corona should be comparable to, or exceed, the local Keplerian speed.

Now let us suppose that the magnetic stress term is systematically negative, and that owing to the factor of the Keplerian velocity, the term $- (4\pi)^{-1} \langle B_\phi^\prime B_z^\prime \rangle \vK $, is significantly greater than $\bar \rho^+ \langle \vA^2 \rangle^+ \tilde v_z^+$. Then, the condition for the energy flux to be positive is:
\begin{equation}
\frac {-\langle B^\prime_\phi B^\prime_z \rangle}{4 \pi} \, \vK > \left( \frac {1}{2} \vK^2 + \phiG \right)
\bar \rho^+ \tilde v_z^+
\end{equation}
and since $\phiG = -\vK^2$ and using equation~(\ref{e:dM}) for $d \Mdotw /dr $, this condition becomes:
\begin{equation}
\left| \frac {d \Mdotw}{dr} \right| \ltapprox  8 \pi r \frac {- \langle B^\prime_\phi B^\prime_z \rangle^+}
{4 \pi \vK}
\label{e:wind_condn2}
\end{equation}
that is, the existence of a wind, in this case implies an upper limit on the wind mass-loss rate.

A more detailed investigation of the conditions for the initiation of disk winds and the corresponding mass-loss limits is deferred to future work. We note, however, the results of \citet{meier97} and \citet{meier99}, pertaining to the case of a nett poloidal field, showing a change in character of disk winds, from loosely collimated to jet--like when the Alfven speed increases from below $\vK$ to above that parameter.
Their results and the above order of magnitude estimate suggest that the condition~(\ref{e:wind_condn1}) may be an interesting critical value. Note, however, that equation~(\ref{e:wind_condn2}) is potentially a weaker condition on the magnetic field, and can be expressed as:
\begin{equation}
\frac {\langle \vA^2 \rangle^+}{\vK^2} \, \left( -\frac {B^\prime_z}{B^\prime_\phi} \right)
\gtapprox \frac{1}{2} \frac{\tilde v_z^+}  {\vK } 
\sim \frac {h_{\rm av}}{r}  \frac {\tilde v_z^+}{c_0} 
\label{e:wind_condn3}
\end{equation}
Given that the vertical wind velocity (at the base of the wind) is likely to be no greater than the sound speed and since $h_{\rm av} /r << 1$ for a non-advective disk, the right hand side of this inequality is much less than unity. Thus, even allowing for the fact that the vertical component of the magnetic field may be significantly less than the magnitude of the magnetic field, the coronal Alfven speed implied by equation~(\ref{e:wind_condn3}) should be considerably less than the local Keplerian speed.  Notwithstanding these estimates and the work by Meier and colleagues, it will be of interest to assess the validity of these estimates numerically especially in the case of a zero nett magnetic flux.

Satisfaction of the above conditions (but especially (\ref{e:wind_condn3})) is possible in principle, even if the magnetic field in the interior of the disk is weak enough for the MRI instability to apply.
If magnetic field is driven into the corona by buoyancy and the magnetic field does not decrease as rapidly as the square root of the density then the Alfv\'{e}n speed in the corona could become quite high. This is, in fact, a feature of the \citet{millstone00} simulations which therefore support the feasibility of a disk wind, in the case of zero nett magnetic flux, even though a strong wind was not produced in their work.

\subsection{Angular momentum}
\label{s:ang_mom}

The azimuthal component of eqn.~(\ref{e:motion_mean}) is:
\begin{equation}
\frac{1}{r^2} \frac{\partial}{\partial r} \left(
r^2 \bar \rho \tilde v_r \tilde v_\phi \right)
+ \frac {\partial}{\partial z} \left( \bar \rho \tilde v_\phi \tilde 
v_z \right) = \frac{1}{r^2}
\frac{\partial}{\partial r}\left( r^2 \langle t_{r \phi} \rangle  \right)  
+ \frac{\partial \langle t_{\phi z} \rangle}{\partial z} \qquad .
\label{e:azimuthal}
\end{equation}
Let us define the vertically integrated $r\phi$ stress by
\begin{equation}
T_{r \phi} = \int_{-h}^{+h} t_{r \phi} \> \partial z
\label{e:int_stress}
\end{equation}

Integrating equation~(\ref{e:azimuthal}) over $z$ and applying mass continuity (\ref{e:dM})
gives
\begin{equation}
\dr \left[ \Mdota \tilde v_\phi r + 2\pi r^2 
T_{r\phi} \right]
=\tilde v_\phi r \frac{{\rm d}\Mdota}{{\rm d}r}
- 4\pi r^2 \langle t_{\phi z} \rangle^+   
\label{e:angmom}
\end{equation}
The terms on the the left hand side of this  equation describe radial transport
of angular momentum due to the mean radial inflow and turbulent MHD stresses,
while the terms on the right hand side describe vertical angular momentum transport
due to mass loss in a wind and turbulent stresses on the disk surface.
Although we do not explicitly consider the effects of a nonzero mean magnetic
field here, we note that equation  (\ref{e:angmom}) can be straightforwardly generalized to
include electromagnetic torque terms analogous to the turbulent MHD stress terms
involving $T_{r\phi}$ and $\langle t_{\phi z} \rangle$.

Radially integrating (\ref{e:angmom}) gives
\begin{equation}
\Mdota  \tilde v_\phi   r 
- \Mdota (\ro) \tilde v_\phi (\ro)  \ro
+ 2\pi r^2 T_{r \phi} - 2\pi \ro^2 T_{r \phi}(\ro)  =
 \int_{\ro}^{r} \> {\rm d}r  \left[ \, \tilde v_\phi r \frac{{\rm d}\Mdota}{{\rm d}r}
- 4 \pi r^2  \langle t_{\phi z}\rangle^+ \,  \right] \qquad ,
\label{e:angmom_int0}
\end{equation}
and using the integrated continuity equation, $\Mdota + \Mdotw = \dot M$,
this can be conveniently expressed in the form:
\begin{equation}
\Mdot \tilde v_\phi  r \zeta
+ 2\pi r^2 T_{r \phi} - 2\pi \ro^2 T_{r \phi}(\ro)  =
 \int_{\ro}^{r} \> {\rm d}r  \left[ \, \Mdotw \dr (\tilde v_\phi r )
- 4 \pi r^2  \langle t_{\phi z}\rangle^+ \,  \right] \qquad ,
\label{e:angmom_int}
\end{equation}
where
\begin{equation}
\zeta (r) = 1 - \frac{\tilde v_\phi (\ro) \ro}{\tilde v_\phi (r) r} \qquad .
\label{e:zeta}
\end{equation}
Note that we do not assume the $r\phi$ stress is zero at $r=\ro$.
While the assumption $T_{r\phi}(\ro) \simeq 0$ is often made in conventional
disk models, its justification rests on the premise (in $\alpha$-disk models)
that the stresses scale as the disk gas pressure, which falls off dramatically
as the matter passes accross the marginally stable orbit.
When the stresses are turbulent, on the other hand, this assumption is no
longer valid. Indeed, numerical simulations  \citep{hawkrol02} show that turbulent stresses
persist at and inside $\ro$.
As can be inferred from (\ref{e:angmom_int}), a nonzero
$r\phi$ stress at $\ro$ requires an increased torque at large $r$ 
to transport the same amount of angular momentum over the same radial distance.
The fact that $T_{r\phi} (\ro)$ does not generally vanish also
has important implications for energy balance in a turbulent disk, since this stress does work on the rotating disk and the resultant turbulent energy can
be dissipated at a comparable rate, thereby providing an additional source
of internal heating at $\ro$.
This is demonstrated explicitly in \S~\ref{s:energy_budget} below.

An important property of MRI-driven turbulence is that the $r\phi$ stress
has the same sign as $\partial \Omega /\partial r$ \citep{balbhaw98} and
thus, even in the absence of a nett vertical angular momentum flux, the
$T_{r\phi}$ stress alone can facilitate the outward transport of angular
momentum required for accretion to proceed.
Whether accretion proceeds at an interesting rate, however, is another issue
deserving separate attention.
Consider a Keplerian disk in which the mass accretion rate varies as
$\Mdota (r) \propto r^p$, where $0 < p < 1$ is the mass loss index used,
for example, by \citet{blandbegel99} \citep[see also][]{becker01}.
Substituting into the integrated angular momentum equation (\ref{e:angmom_int0})
yields
\begin{equation}
\frac{1}{2p + 1} \left[
1 - \left( \frac{r}{\ro} \right)^{-(p+\case{1}{2})} \right] \simeq
\frac{G_{r\phi}(r)}{\Mdota \vK r}
\left[ 1 - \frac{G_{r\phi} (\ro)}{G_{r\phi} (r)} \right]
+ \frac{G_{\phi z}(r)}{\Mdota \vK r}
\left[ 1 - \frac{G_{\phi z} (\ro)}{G_{\phi z} (r)} \right]
\qquad ,
\label{e:G}
\end{equation}
where $G_{r\phi}(r) = 2\pi r^2 (-T_{r\phi})$  and
$G_{\phi z}(r) = -\int_{\infty}^r 4\pi r^2 {\rm d}r
\langle t_{\phi z} \rangle^+$ are the
torques associated with the turbulent MHD stresses.
This implies that if $G_{r\phi} , G_{\phi z}\propto r^{-(p+\case{1}{2})}$, then
\begin{equation}
\Mdota \vK r \simeq \left( \, G_{r\phi} + G_{\phi z} \, \right)\,(2p + 1)
\end{equation}
which in turn implies that when vertical transport of angular momentum
is taken into account, the mass accretion rate is enhanced by a factor
$(1 + G_{\phi z}/G_{r\phi})(2p+1)$.
Conversely, when the vertical angular momentum flux is neglected, then
$T_{r\phi}$ scales as $\tilde v_r \vK$, a result obtained in numerical
models \citep{hawkrol02}.
This result can be used to estimate the minimum mass accretion rate at
$\ro$, where the MRI-driven turbulent stresses saturate at quasi-thermal
levels (i.e. $t_{r\phi}(\ro) \simeq \bar \rho (\ro) c_{\rm s}^2 (\ro)$).
Using $\Mdota = 2\pi r \Sigma ( -\tilde v_r )$ gives
\begin{equation}
\Mdota (r) = \varepsilon \, \tauT \, \frac{\ro}{\rg}
\frac{(-\tilde v_r)}{c} \, \dot M_{\rm Edd} \qquad ,
\end{equation}
where $\tauT = \int_0^h (\sigmaT / m_{\rm p}) \bar \rho {\rm d}z 
= \case{1}{2} \Sigma \sigmaT /m_{\rm p}$ is the
Thomson optical depth of the disk over its vertical scaleheight
and $M_{\rm Edd} = 4\pi GMm_{\rm p}/(\varepsilon \sigmaT c)
\simeq 0.3 \varepsilon_{0.1}^{-1} M_7 M_\odot\,{\rm yr}^{-1}$
is the Eddington mass accretion rate for a conversion efficiency
$\varepsilon = 0.1\varepsilon_{0.1}^{-1}$.
The condition $-\tilde v_r (\ro) \simeq - t_{r\phi}(\ro)/\vK(\ro)
\simeq c_{\rm s}^2(\ro) / \vK (\ro)$ satisfied when angular momentum
is solely transported by the turbulent $r\phi$ stress then implies a
dimensionless mass accretion rate
\begin{equation}
\dot m_{\rm a} (\ro) \equiv \frac{\Mdota  (\ro)}{\dot M_{\rm Edd}}
\gtapprox \, 3 \times 10^{-3} \, \varepsilon_{0.1} \, \tau_5 (\ro) \,
r_{\rm i,10}^{3/4} \, T_5(\ro) \qquad ,
\label{e:mdota_ri}
\end{equation}
where $T_5 (\ro) = T(\ro)/10^5\,$K and $\tau_5(\ro) = \tauT (\ro) / 10^5$ are
approximately unity for a typical AGN disk.
For example, accreting gas will
come into thermal and radiative equilibrium and cool to the local blackbody
temperature 
$T_{\rm bb} \simeq 2 \times 10^5(\Ld/L_{\rm Edd})^{1/4}
(\ro/10\rg)^{-1/2}M_7^{-1/4}\,$K
when the number density reaches
$\bar n_0 (\ro) \sim 10^{18}\,{\rm cm}^{-3}$
\citep*[e.g.][]{kuncic97}
, which implies
$\tauT (\ro) \simeq 10^5$ at
$\ro = 10r_{\rm g} \simeq 1.5 \times 10^{13}M_7\,$cm for
a thin disk with $h(\ro) \simeq 10^{-2}\ro$.
An immediate implication of this result is that since $\Mdota$ is constant when
there is no wind, then unless the vertical Thomson depths are unusually high,
{\em moderate to high mass accretion rates in geometrically-thin,
optically-thick disks cannot be attributed to weak-field MHD turbulence alone}.
Indeed, substantially sub-Eddington mass accretion rates are a characteristic
property of numerical models where angular momentum is transported solely
by $t_{r\phi}$ \citep[see e.g.][]{stonepring01,hawley01,hawbalb02}.
This limitation arises from the weak-field nature of the MRI, which becomes
ineffective once the turbulence reaches thermal levels.
This  constrains the maximum inflow speeds that can be attained in
the accretion flow such that $|\tilde v_r|/c_{\rm s} \simeq c_{\rm s}/\vK$.
Therefore, unless there is an additional instability capable of
driving MHD turbulence from the weak to the strong regime, angular momentum
transport in disks with moderate to high mass accretion rates must be largely
attributed to a combination of vertical mass loss and mean-field torques
(including large-scale MHD outflows, which we have not explicitly included here).

Note, however, that the role of the $G_{\phi z}$ torque in vertical angular
momentum transport remains unclear.
In princple,  $G_{\phi z}$ could be larger than  $G_{r\phi}$ by as much
as $r/h$, in which case this surface torque could play a key role in the
energetics of thin disks.
Consider the condition (\ref{e:angmom}) for angular momentum conservation.
For a Keplerian disk, this implies
\begin{equation}
\Mdota \frac {\rm d}{{\rm d}r} \left(r \vK \right) + \frac {\rm d}{{\rm d}r} \left( 2 \pi r^2 T_{r\phi} \right) =
- 4 \pi r^2 \langle t_{\phi z} \rangle^+
\label{e:angmom2}
\end{equation}
It is then straightforward to show that the $\phi z$ stress has an important effect on angular momentum transport when
\begin{equation}
\frac {- 8 \pi r^2 \langle t_{\phi z} \rangle^+}{\Mdota \vK} \sim 1
\label{e:az_vert_condn}
\end{equation}
and that the $\phi z$ stress is important relative to the $r \phi$ stress when
\begin{equation}
\langle t _{\phi z} \rangle^+ \sim \left( \frac {h_{\rm av}}{r} \right) \, \langle t_{r \phi} \rangle_0 
\label{e:tphiz_condn}
\end{equation}
This extraordinary result reflects  the fact that $\langle t _{\phi z} \rangle^+$ acts over the surface of the disk but $\langle t_{r \phi} \rangle_0$ acts only over the height. The simulations reported by \citet{millstone00} produce a volume-averaged $\langle t _{\phi z} \rangle^+ \sim 0.02 \times \langle t_{r \phi} \rangle_0$.
Even this low a stress is capable of physically interesting effects, as we argue in the following section
when we consider the implications of mass and momentum transport for the energetics of
turbulent accretion disks. 

\section{The disk    energy budget}
\label{s:energy_budget}

We now apply the above deductions from the momentum balance in a magnetized,
turbulent accretion disk to calculate the power generated in the disk and emerging
from the disk surface as both radiation and mechanical plus non--radiant
electromagnetic energy.
Dissipation of the latter two components may generate further emissivity in a corona
and/or outflow.
Our basis for this calculation is the total energy equation~(\ref{e:energy_mean}).
Usually in expositions of accretion disk theory, the internal energy equation is used,
since this  explicitly identifies energy exchange processes, in particular, the transfer
of free energy from the Keplerian shear in the bulk flow to the stresses which then
dissipate that energy in the disk.
In a turbulent MHD disk, this energy transfer is largely attributed to the turbulent
energy production term $\langle t_{ij} \rangle \tilde s_{ij}
\simeq \langle t_{r\phi} \rangle \tilde s_{r \phi}$, that is related to the rate
of viscous dissipation via a turbulent cascade and the rate of bulk heating of the
fluid via various additional radial and vertical transport terms.
However, the internal energy equation for a turbulent, MHD accretion disk involves
many intermediate terms that are cumbersome.
The total energy equation, on the other hand, has the advantage that
all terms are in total fluxes, which can be integrated straightforwardly
to obtain a conservation equation in terms of the nett power in each of the
available energy fluxes.
We therefore prefer to use the total energy equation, which more elegantly describes
the contributions of all terms to the overall energy balance in a disk.
 
\subsection{Order of magnitude estimates}
\label{s:estimates}
The total energy flux contains a number of terms and in estimating the relative
importance of  these, it is useful to deduce order of magnitude estimates of the various
components of velocity.
These are the mean velocity components, $\tilde v_r$, $v_\phi = (GM/r)^{1/2}$ and
$\tilde v_z$, and the corresponding fluctuating components $v^\prime_r$, $v_\phi^\prime$
and $v_z^\prime$.

The component $\tilde v_r$ is the radial inflow speed associated with the mass accretion
rate {\it viz.} $\tilde v_r = - \Mdota/2\pi r\Sigma$.
An estimate of this quantity depends upon which stress dominates
equation~(\ref{e:angmom2}).
If the $r\phi$ stresses dominate, then $|\tilde v_r \tilde v_\phi | \sim
|T_{r\phi}|/\Sigma$, implying that 
\begin{equation}
\frac {|\tilde v_r|}{\cs} \sim \frac{\cs}{\vK}
\frac{|\langle t_{r\phi} \rangle|}{\bar \rho_0 \cs^2} 
\qquad .
\label{e:vr_rphi}
\end{equation}
If the $\phi z$ stresses dominate angular momentum transport, then approximating the
radial derivative in equation~(\ref{e:angmom}) by division by $r$, we obtain
\begin{equation}
\frac {|\tilde v_r|}{\cs} \sim \frac{r}{h} \frac{\cs}{\vK}
\frac{|\langle t_{\phi z} \rangle^+|}{\bar \rho_0 \cs^2} \qquad .
\label{e:vr_phiz}
\end{equation}

If we take $|\langle t_{r \phi} \rangle | \sim \rho_0 v^{\prime 2}$ (where $v^\prime $
is the magnitude of the turbulent velocity), then since the turbulence is weak
($v^{\prime 2} \ltapprox \cs^2$), the first estimate of $\tilde v_r$
based on the $r\phi$ stresses implies $|\tilde v_r|/\cs \ll 1$  for an optically-thick
disk (see also the estimates in \S~\ref{s:ang_mom} above).
If the $\langle t_{\phi z} \rangle$ stresses are of similar magnitude to
$\langle t_{r \phi} \rangle$ then $\tilde v_r$ is larger by a factor of $r/h$ when
angular momentum transport by $\langle t_{\phi z} \rangle^+$ is important.
Thus, if the disk is sufficiently thin that $h_{\rm av}/r \ltapprox \langle t_{\phi z}\rangle^+/
\langle t_{r\phi} \rangle$, then the $\phi z$ stresses could increase the radial Mach
number $\tilde v_r/\cs$ to a value above $\cs /\vK$. So far, simulations of turbulent accretion disks \citep[e.g.][]{millstone00} estimate these
stresses to be approximately an order of magnitude smaller than the
$\langle t_{r\phi} \rangle$, since the MRI preferentially amplifies the $r$ and $\phi$
components of the fluctuating fields.

The conclusion from these estimates is that the radial Mach number is likely to
be small under most conditions.
This means that most terms in the total energy equation involving $\tilde v_r$ can be
neglected, except those involving multiplication by a large quantity such as
$\tilde v_\phi \simeq \vK$ or the gravitational potential. 

As far as the relative values of kinetic and magnetic stresses are concerned, we rely to some extent,
though not exclusively on numerical results, which show that the
turbulent magnetic stresses are a factor of a few higher than the Reynolds stresses.

The fluctuating velocity components $v_r^\prime$ and $v_\phi^\prime$ are less than the sound speed, so that we neglect the turbulent energy
flux terms related to these quantities.
However, vertical gradients in turbulent energy flux terms involving $v_z^\prime$
should in general be retained since, in addition to being affected by the MRI, this
component of the fluctuating velocity can also be enhanced by the Parker instability
and the total rate of vertical energy transport can be nonnegligible in a thin disk.
Turbulent perturbations in a magnetized disk become unstable to the gravitational modes
of the Parker instability as matter drains down slightly elevated flux tubes, thereby
enhancing their buoyant rise to the disk surface.
The buoyant velocity field generated in this way can legitimately be considered as a
turbulent velocity since regions of rising underdense flux tubes are balanced by falling,
overdense regions and the nett  mass-averaged velocity is zero.
We estimate the buoyant velocity of such a flux tube by balancing the buoyant force per
unit length corresponding to a density deficit of $-\delta \rho$ with the drag force
$\propto C_{\rm d} {v_z^\prime}^2 $ where $C_{\rm d} \sim 1$ is the drag coefficient.
Thus we estimate the vertical buoyant velocity of a tube of radius $R_{\rm tube}$ at a
height $z$ to be \citep[e.g.][]{parker79}:
\begin{eqnarray}
v_z^\prime & \approx & \left( \frac {\pi}{C_{\rm d}} \right)^{1/2}
\, \left( \frac {|\delta \rho|}{\rho} \right)^{1/2}
\,
\left( \frac {z R_{\rm tube}}{h^2} \right)^{1/2} \, \vK \,
\left( \frac {h_{\rm av}}{r} \right)
\nonumber \\
& \ltapprox & \left( \frac {\pi}{C_{\rm d}} \right)^{1/2} \,
\left( \frac {|\delta \rho|}{\rho} \right)^{1/2}
\,
\left( \frac {z R_{\rm tube}}{h^2} \right)^{1/2} \, c_0
\label{e:vb}
\end{eqnarray}
This velocity depends upon the density contrast generated, the height of the tube above
the midplane and the tube radius, all of which are uncertain.
For modest values of $\delta \rho/\rho$ and $R_{\rm tube}/h_{\rm av}$
the buoyant velocity can be an appreciable fraction of $c_0$ but it unlikely to be greater than $c_0$.
The \citet{millstone00} simulations, show an average kinetic energy associated with
$v_z^\prime$ comparable to that associated with $v_r^\prime$ and $v_\phi^\prime$,
consistent with this.

As we have indicated earlier, the vertical mean flow velocity at the base of the corona, $\tilde v_z^+$,  that could arise from an imbalance between the thermal and radiation pressures and the local
gravitational field, cannot be estimated without a specific model. Nevertheless, the order of magnitude argument advanced in \S~\ref{s:disc_eqns} suggests that it may be as large as the disk sound speed. Lacking any detailed  knowledge of the magnitude of this velocity component, we retain terms in the vertical energy flux associated with it.

In applying the total energy equation, we also neglect the viscous and resistive terms in the energy flux, $-\langle \tv_{ij} v_{ij}^\prime \rangle + \langle  \eta \tB_{ij} \rangle_{,j} $.
As we noted in \S~\ref{s:mean_field}, these terms could be locally important in,
for instance, shocks or reconnection regions but when the energy equation is
integrated over a large volume, their contribution to the surface integral is minor
\citep[see e.g.][ for a discussion on resistive heating in turbulent disks]{hawbalb02}.
Moreover, it is unlikely that the almost singular surfaces on which these terms are important would intersect substantially with the annular surfaces, $r<r<r+{\rm d}r$, $-h < z < h$,  over which we integrate the total energy equation. Also, neither term is important in the high wave number end of a turbulent cascade since they involve one less spatial differentiation than the Joule heating and viscous dissipation terms that appear in say, the internal energy equation. Those terms are locally important but they are not required in the integration of the total energy equation.

\subsection{Application of the total energy equation}
\label{s:etot}

Let us now calculate the nett power available from each of the energy flux terms
in the statistically-averaged  equation for total energy conservation, given
by (\ref{e:energy_mean}), taking into account the order of magnitude estimates
deduced in the preceding section.  
For an axisymmetric, steady-state disk:
\begin{equation}
\label{e:divF}
\frac{1}{r}\frac{\partial}{\partial r} (r \langle \FE_r \rangle ) \, + \,
\frac{\partial \langle \FE_z \rangle}{\partial z} \, = \, 0  \qquad ,
\end{equation}
where $\langle \FE_r \rangle$ and $\langle \FE_z \rangle$ are the radial and
vertical components of the total mean energy flux $\langle \FE_i \rangle$,
defined by (\ref{e:Fe2}).
Integration over the disk height yields:
\begin{equation}
\dr \int_{-h}^{+h} 2 \pi r \langle \FE_r \rangle  \> \dz
\, + \, 4 \pi r {\langle \FE_z \rangle}^+  \, = \, 0
\label{e:disk_energy}
\end{equation}
and this equation is the basis of the calculation of both the differential and total
radiative energy flux from the disk.

The dominant terms in the radial component of the total energy flux give
\begin{equation}
\FE_r \simeq \left(  \frac{1}{2} \tilde v^2 + \phiG  \right) \bar \rho \tilde v_r
- \langle t_{r \phi} \rangle \tilde v_\phi
+ \langle F_r \rangle  +  \langle  Q_r \rangle
\qquad .
\end{equation}
Upon vertical integration over a thin disk, followed by differentiation with respect to radius the last two terms become ${\cal O} (h_{\rm av}/r)$ compared to analogous terms in the vertical energy flux, implying
\begin{equation}
\int_{-h}^{+h} 2 \pi r \FE_r \> \dz \simeq 
\left( \frac {1}{2} \tilde v_\phi^2 + \phiG  \right)\Mdota
- 2 \pi r T_{r \phi}\tilde v_\phi 
\label{e:fr}
\end{equation}
where $T_{r \phi}$ is the vertically-integrated $r \phi$ stress defined by
(\ref{e:int_stress}).
The term $- 2\pi r T_{r \phi}\tilde v_\phi \simeq - 2\pi r T_{r \phi}\vK$
can be eliminated using equation
(\ref{e:angmom_int}) for angular momentum conservation:
\begin{equation}
- 2\pi r T_{r \phi}\vK = \Mdot \vK^2 \zetaK^{(1)}(r) - 2 \pi \ro^2 T_{r \phi}(\ro) \OmegaK(r) 
- \OmegaK \int_{\ro}^r \left[ \Mdotw (r^\prime) \frac{\rm d}{{\rm d}r^\prime}
(r^\prime \vK(r^\prime)) 
-4 \pi r^{\prime 2} \langle t_{\phi z} (r^\prime) \rangle^+   \right] \> {\rm d}r^\prime
\label{e:torque}
\end{equation}
where
\begin{equation}
\zetaK^{(1)}(r) = 1 - (\ro/r)^{1/2} \qquad .
\end{equation}

The vertical component of the energy flux is:
\begin{eqnarray}
\langle \FE_z \rangle^+ &=&
\left[ \frac {1}{2} \bar \rho^+ \tilde v^{+2} + \bar \rho^+ \phiG + \bar \rho^+ \tilde h^+ 
+ \langle \uK \rangle^+ + \langle \uB \rangle^+ \right] \tilde v_z^+ 
+ \langle \rho h^\prime v_z^\prime \rangle^+ + \langle \uK v_z^\prime \rangle^+
+ \langle \uB v^\prime_z \rangle^+   \nonumber \\
&&
- \langle t_{\phi z}  \rangle^+  \tilde v_\phi - \langle t_{zz}  \rangle^+  \tilde v_z^+
- \langle t_{zz} v^\prime_z \rangle^+ + \langle F_z \rangle + \langle Q_z \rangle
\qquad .
\end{eqnarray}
There are some terms in this expression such as the first two terms representing the
vertical flux of kinetic plus gravitational energy in a wind and the term
$\langle t_{\phi z} \rangle^+ v_\phi^+$, that we should obviously keep.
There are also some terms in this equation that we can immediately discard.
The term $\bar \rho^+ \tilde h^+ \tilde v_z^+ = (\bar u^+ + \bar p^+) \tilde v_z^+$
is small at the photosphere-corona boundary.
For similar reasons we can also neglect the turbulent enthalpy flux term
$\langle \rho h^\prime v^\prime_z \rangle $.
In view of the order of magnitude estimates considered in \S~\ref{s:estimates},
the remaining terms may be comparable.
To simplify the final expression, we combine the remaining terms involving a
vertical velocity component into an energy flux term $\langle \psi_z \rangle^+$,
defined by
\begin{equation}
\langle \psi_z \rangle^+ = 
\left( \langle  \uK + \uB  \rangle^+ - \langle t_{zz} \rangle^+  \right) \tilde v_z^+
+ \langle \left( \uK + \uB  - t_{zz} \right) v^\prime_z \rangle^+
-\langle t_{\phi z} v^\prime_\phi \rangle^+  \qquad .
\end{equation}
With these simplifications, the vertical component of the total energy flux
reduces to
\begin{equation}
4 \pi r \langle {\FE_z} \rangle^+ \simeq
4 \pi r \left( \frac {1}{2} \vK^2 + \phiG \right) \bar \rho^+ \tilde v_z^+ \, 
- \, 4 \pi r \langle t_{\phi z} \rangle^+ \tilde v_\phi 
\, + \, 4 \pi r \left( \langle \psi_z \rangle^+
+  \langle F_z  \rangle^+ + \langle Q_z \rangle^+ \right) \qquad .
\label{e:fz}
\end{equation}

Subsitution of equations~(\ref{e:fr}) and (\ref{e:fz}) into the disk total energy
equation~(\ref{e:disk_energy}), utlizing equation~(\ref{e:torque}) gives the following
for the radiative flux from both sides of the disk (after some algebra):
\begin{eqnarray}
\label{e:Ftot}
4 \pi r \langle F \rangle^+ &=& \frac{3}{2}\frac {G M \Mdot}{r^2}  \zetaK^{(1)}(r)
- 3 \pi \ro^2 T_{r \phi}(\ro) \frac{\Omega_K(r)}{r}
\nonumber \\
&& - \frac{3}{2}\frac{\Omega_K(r)}{r} \int_{\ro}^r 
 \left[ \frac{1}{2}\Mdotw(r^\prime) \vK(r^\prime) 
- 4 \pi {r^\prime}^2 \langle t_{\phi z} (r^\prime) \rangle^+  \right]
\> {\rm d}r^\prime \nonumber \\
&& - 4 \pi r \langle \psi_z \rangle^+ - 4 \pi r \langle Q_z \rangle^+
\qquad .
\end{eqnarray}
The first term on the right of this expression represents the binding energy flux
associated with the nett mass flux and is familiar from standard disk theory
\citep[c.f.][]{SS73}.
The second term represents a correction resulting from a non-vanishing stress at the
inner radius.
The third term is related to the flux of kinetic and gravitational energy in the wind; this term is discussed further below.
The fourth term represents the effect of the $\phi z$ turbulent stresses and its
magnetic component is intimately associated with the Poynting flux into the corona;
this term is also discussed further below.
The fifth term, $- \langle \psi_z \rangle^+$, represents another component of the wind power associated with the advective part of the Poynting flux and advection of other turbulent quantities.
The fifth term, $-\langle Q_z \rangle $,  represents the effect of an external heat
flux arising from above the disk. Since $\langle Q_z \rangle < 0$, this heating enhances the radiative flux from the disk.

The above form of the radiative flux is convenient for integration since it involves the (constant) total mass accretion rate, $\Mdot$ in the first term and the factor $\zetaK^{(1)}(r)$ has a simple analytical form. Nevertheless, in order to facilitate comparison of the various terms, the third term may be integrated by parts and partially combined with the first term to give:
\begin{eqnarray}
\label{e:Ftot2}
4 \pi r \langle F_z \rangle^+ &=& \frac{3}{2}\frac {G M \Mdota}{r^2}  \zetaK^{(2)}(r)
- 3 \pi \ro^2 T_{r \phi}(\ro) \frac{\Omega_K(r)}{r}
\nonumber \\
&& - \frac{3}{2}\frac{\Omega_K(r)}{r} \int_{\ro}^r 4 \pi r^{\prime 2}
\left[ \bar \rho^+ \tilde v_z^+ \vK(r^\prime)
- \langle t_{\phi z} (r^\prime) \rangle^+  \right]
\> {\rm d}r^\prime \nonumber \\
&& - 4 \pi r \langle \psi_z \rangle^+ - 4 \pi r \langle Q_z \rangle^+
\qquad .
\end{eqnarray}
where
\begin{equation}
\zetaK^{(2)}(r) = 1 - \left( \frac {\ro}{r} \right)^{1/2} \frac {\Mdota(\ro)}{\Mdota(r)}
\end{equation}
The form~(\ref{e:Ftot2}) of the radiative flux from the disk more clearly shows the effect of the local accretion rate in the first term and the effect of the angular momentum transported by the wind and the $\phi z$ stress in the third and fourth terms respectively.
 
An expression for the {\em total} disk radiative luminosity, $\Ld$, may be obtained by integrating each term of equation~(\ref{e:Ftot}) over all disk radii, from $r = \ro$ to $r = \infty$. This also helps to elucidate the significance of the various terms in the expressions for the radiative power from the disk. The result is: 
\begin{eqnarray}
\label{e:Ld}
\Ld &=& \frac {1}{2} \frac {GM\Mdot}{\ro}
- 2 \pi \ro^2 T_{r \phi}(\ro) \OmegaK(\ro)  
- \frac{1}{2} \frac{GM\Mdotw (\ro)}{\ro}
- \int_{\ro}^\infty \, \frac{1}{2}\frac{GM}{r} \frac{{\rm d}\Mdotw}{{\rm d}r}{\rm d}r
\nonumber \\
&&
+ \int_{\ro}^\infty 4 \pi r \langle t_{\phi z}(r) \rangle^+\vK (r)  \> {\rm d}r 
  - \int_{\ro}^\infty 4 \pi r \,  \langle \psi_z \rangle^+ \> {\rm d}r
  - \int_{\ro}^\infty 4 \pi r \, \langle Q \rangle^+ \>  {\rm d}r
\end{eqnarray}
The first and third terms combine to give the familiar expression for the accretion power:
\begin{equation}
\Pa = \frac {1}{2} \frac {GM\Mdota (\ro)}{\ro} \qquad .
\end{equation}
In conventional accretion disks, this is the dominant term responsible for the disk
luminosity.
In the generalized formalism developed here, however, we take into account disk solutions
in which $\Mdota$ decreases towards small $r$ owing to the vertical loss of mass.
Thus, for disks with strong wind mass loss, $\Mdota (\ro)$ is a fraction of the nett
mass flux $\Mdot$ and $\Pa$ is not necessarily the dominant term.
 
In considering the effect of a nonvanishing stress at $\ro$, we introduce the parameter
\begin{eqnarray}
\kappa &=& \frac {\hbox{Outward angular momentum flux due to turbulent stresses at } r=\ro}
{\hbox{Inward angular momentum flux due to  accretion at }r=\ro} 
\nonumber \\
& = & \frac {- 2 \pi \ro^2 T_{r \phi}(\ro) }{\Mdota (\ro) \ro^2 \OmegaK(\ro)}
= \frac{T_{r\phi}(\ro)}{\Sigma (\ro) \tilde v_r (\ro) \vK (\ro)}
\end{eqnarray}
Note that since  $T_{r \phi}(\ro) < 0$, consistent with the stresses defined elsewhere
in the disk, this parameter is positive.

We now combine the fourth, fifth, and sixth terms on the right hand side of
(\ref{e:Ld}) into a single term describing the total power removed from the
disk by a wind:
\begin{equation}
\Pw = - \int_{\ro}^\infty \, \frac{1}{2}\frac{GM}{r} \, 4 \pi r \bar \rho^+ \tilde v_z^+ 
\> {\rm d}r
- \int_{\ro}^\infty 4 \pi r \langle t_{\phi z}(r) \rangle^+\vK (r)  \> {\rm d}r 
+ \int_{\ro}^\infty 4 \pi r \,  \langle \psi_z \rangle^+ \> {\rm d}r \qquad .
\end{equation}
The leading term in the wind power is negative since the disk material is highly bound; the only way in which the wind power can be positive is for the succeeding terms in $\Pw$ to be positive. (See the discussion concerning winds below equation~(\ref{e:vz})). 

We define the nett heating rate of the disk due to irradiation and heat conduction from an external corona by
\begin{equation}
\PQ = - \int_{\ro}^\infty 4 \pi r \, \langle Q \rangle^+ \>  {\rm d}r  \geq 0
\end{equation}

We can now write the total disk luminosity as the following sum:
\begin{equation}
\Ld =(1+ 2 \kappa) \Pa - \Pw + \PQ \qquad .
\end{equation}
When vertical energy transport
from the disk (i.e. $\Pw$) is important, the disk luminosity is reduced as a result of the conservation of energy.
Realistically, some of  the total power $\Pw$ removed from the disk is available to
heat the corona and potentially, the most  important contribution to this term is:
\begin{equation}
P_{\phi z} = - \int_{\ro}^\infty 4 \pi r \langle t_{\phi z} \rangle^+ \vK  \> {\rm d}r
\qquad ,
\label{e:Pphiz}
\end{equation}
which represents the integral of part of the Poynting flux over both surfaces of the disk (see equation~(\ref{e:poynting})).
To evaluate the importance of this term we compare the integrand of (\ref{e:Pphiz}) to the leading term in equation~(\ref{e:Ftot2}) for the disk energy flux. This comparison shows that $P_{\phi z}$ is an important component of the disk energy budget when
\begin{equation}
\frac {8 \pi r^2}{3} \frac {-\langle B^\prime_\phi B^\prime_z \rangle^+}{4 \pi \Mdota \vK} \sim 1
\label{e:Pphiz_condn}
\end{equation}
We showed in \S~\ref{s:disc_eqns} (see equation~(\ref{e:az_vert_condn})) that
$\langle t_{\phi z}\rangle$ has an appreciable effect on the accretion process when
\begin{equation}
\frac {- 8 \pi r^2  \langle t_{\phi z} \rangle^+}{\Mdota  \vK} \sim 1
\qquad .
\end{equation}
Hence, when this is the case, the power $P_{\phi z}$ is comparable to $\Pa$, that is, when the
$\phi z$ stresses are important for angular momentum transport, they do significant work on the flow.

Of course, the effect of the $\phi z$ stress would be unimportant if it fluctuated in sign.
However, if the disk has a wind, then one expects $\langle t_{\phi z} \rangle^+< 0$
as a result of inertia causing magnetic field loops to trail the rotation of the disk.
When $\langle t_{\phi z} \rangle^+$ is systematically negative, the disk luminosity is reduced for a given mass accretion rate and a corresponding amount of energy is available to heat the corona.
As we have shown in \S~\ref{s:disc_eqns}, even a weak stress can be dynamically important and the above shows that it can also be energetically important.
Indeed, the importance of $\phi z$ stresses in thin disks is widely appreciated
in accretion-outflow models.
In the model by \citet{pudritz83}, for instance, the entire accretion flow
is attributed to outward angular momentum transport by the torque associated
with these stresses. A major physical difference between our model and that of disk-driven wind models such as that of \citet{pudritz83} and \citet{blandpayne82} is in the mean magnetic flux threading the disk.

We can also express the condition~(\ref{e:Pphiz_condn}) in terms of other disk parameters using, $\Mdota \sim 4 \pi r h_{\rm av} \rho_0 |\tilde v_r|$. For $P_{\rm \phi z}$ to be a significant fraction of the accretion power
\begin{equation}
\frac {-\langle B^\prime_\phi B^\prime_z \rangle}{4 \pi \bar \rho^+ \vK^2} \sim
\frac {\rho_0}{\rho^+} \frac {h_{\rm av}}{r} \frac {\tilde v_r}{\vK}
\sim \frac {\rho_0}{\bar \rho^+} \left( \frac {h_{\rm av}}{r} \right)^2 \, \frac {|\tilde v_r|}{c_0} 
\end{equation}
For representative AGN values $\rho_0/\rho^+ \sim 10^6$, $h_{\rm av}/r \sim 10^{-3}$, this
condition effectively implies (c.f. eqn.~(\ref{e:wind_condn3}))
\begin{equation}
\frac {\vA^2}{\vK^2} \frac{B_z^\prime}{B_\phi^\prime} \sim
 \> \frac {|\tilde v_r|}{c_0}
\end{equation}
If the radial inflow velocity is subsonic, and there is at least a modest vertical field (e.g. $B^\prime_z/B_\phi^\prime \sim 0.1$, say), then this condition may be satisfied if
the coronal Alfv\'{e}n speed is comparable to the Keplerian speed. However, depending upon the details of disk parameters (including $h_{\rm av}/r$ and the radial Mach number), the condition may be satisfied for much lower values of the Alfv\'{en} speed.

At this stage we also consider further the relationship of the power $P_{\phi z}$, to the vertical component of the Poynting flux:
\begin{equation}
\langle  S_z \rangle  =  \langle \uB \rangle \tilde v_z + \langle \uB v^\prime_z \rangle  
- \langle \tB_{rz} \rangle  \tilde v_r - \langle \tB_{\phi z} \rangle  \tilde v_\phi - \langle \tB_{zz} \rangle  \tilde v_z
- \langle \tB_{rz}  v_r ^\prime \rangle
- \langle \tB_{\phi z}  v_\phi ^\prime \rangle
- \langle \tB_{zz}  v_z ^\prime \rangle \qquad .
\end{equation}
The term 
\begin{equation}
\langle S_z^{(1)}  \rangle = -\langle \tB_{\phi z} \rangle  \tilde v_\phi
\end{equation}
in this expression contributes directly to $P_{\phi z}$.
(Other terms in $\langle  S_z \rangle$ are represented in $\psi_z$.)
The power $P_{\phi z}$ represents the integral of this part of the Poynting flux.
It is interesting that the term
$\langle \tB_{\phi z} \rangle  \tilde v_\phi$ does {\em not} appear in the expression
for the {\em local} radiative flux, {\it viz.} equation~(\ref{e:Ftot}).
Such a term is initially present as the above derivation for the local radiative flux shows.
However, it is eliminated by a corresponding opposite term arising from the expression
for $T_{r \phi}$.
What does remain in the expression for the local radiative flux is a more complicated
term involving $\OmegaK$ multiplied by an integral over $\langle \tB_{\phi z} \rangle$. This cause of this is that the angular momentum flow is driven by the total stress whereas the Poynting flux is only related to the magnetic part of that stress as well a the inter-realtionship between the $r\phi$ stress and the $\phi z$ stress in the angular momentum equation~(\ref{e:angmom}).

The term $\int_{\ro}^\infty 4 \pi r \psi_z \> dr$ involves the integration of a number
of terms all of which could be important in the disk energy budget.
Rather than consider them all here, we consider the representative, and probably most important terms:
\begin{eqnarray}
4 \pi r \langle S_z^{(2)}  \rangle &=&
4 \pi r \left[ \langle \uB  -  \tB_{zz} \rangle^+ \tilde v_z^+ \right]
= 4 \pi r \left[ \frac{\langle {B_r^\prime}^2 + {B_\phi^\prime}^2 \rangle^+}{4 \pi}
\tilde v_z^+ \right]  \\
4 \pi r \langle S_z^{(3)} \rangle &=&
4 \pi r   \langle \left(\uB  -   \tB_{zz} \right)  v_z ^\prime \rangle^+   
= 4 \pi r\left[ \frac{\langle ({B_r^\prime}^2 + {B_\phi^\prime}^2 ) v^\prime_z \rangle^+}
{4 \pi}\right]
\end{eqnarray}
The term $\langle S_z^{(2)}  \rangle$ represents the contribution to the Poynting flux
from the wind; the term $\langle S_z^{(3)} \rangle$ represents the contribution to  the
Poynting flux from turbulent diffusion.
Consider the first term,
\begin{equation}
4 \pi r \langle S_z^{(2)}  \rangle = 4 \pi r \bar \rho^+ \tilde v_z^+
\frac {\langle {B^\prime_r}^2 + {B_\phi^\prime}^2 \rangle^+ }
{4 \pi \bar \rho^+}
= - \frac {{\rm d} \Mdotw}{{\rm d}r}
\frac {\langle {B^\prime_r}^2 + {B_\phi^\prime}^2 \rangle^+ }
{4 \pi \bar \rho^+} 
\sim - \langle \vA^2 \rangle^+  \frac {{\rm d} \Mdotw} {{\rm d}r}
\end{equation}
Comparing this with the leading term in equation~(\ref{e:Ftot2}) for $4 \pi r \langle  F_z \rangle$ we have:
\begin{equation}
\frac { \langle \vA^2 \rangle^+  |{\rm d} \Mdotw/ {\rm d}r|}{3 G M \Mdota /2 r^2} 
=   \frac {\langle \vA^2 \rangle^+ }{\vK^2}
\, \frac{r}{\Mdota}\left| \frac{{\rm d}\Mdotw}{{\rm d}r }\right| 
=  \frac {\langle \vA^2 \rangle^+ }{\vK^2}
\, \frac{r}{\Mdota} \frac{{\rm d}\Mdota}{{\rm d}r }
\qquad .
\end{equation}
For the power associated with this term to be important,it would seem that both the wind mass loss
rate should be comparable to the mass accretion rate
($r |{\rm d}\Mdotw/{\rm d}r | \sim \Mdota$)  and the Alfv\'{e}n speed in the corona should be
comparable to the local Keplerian speed. Otherwise, the requirements on each of these factors would probably be excessive. 

Similarly, one can compare the diffusive terms to the same leading term in
$4 \pi r \langle  F_z \rangle$. The diffusive flux is important when
\begin{equation}
\frac { 4 \pi r \langle S_z^{(3)}\rangle }{3 G M \Mdot /2 r^2} 
\sim \frac { \case{\langle B^{\prime 2}\rangle^+}{4 \pi}  v^\prime_z }
{\case{3}{2} \case{GM\Mdota}{r^2}}
\sim 1
\end{equation}
This condition is similar in some respects to the one above referring to the systematic velocity component. However, in this case the turbulent velocity is not related to the wind mass-loss rate and one should allow for the possibility that the turbulent velocity can in principle be larger than the mean wind velocity. Again, utilizing
$\Mdota \sim 4 \pi r h \bar \rho_0 |\tilde v_r|$ in this expression leads to the condition
\begin{equation}
\frac {\langle \vA^2 \rangle^+}{\vK^2} \gtapprox \frac {\bar \rho_0}{\bar \rho^+}
\frac {h_{\rm av}}{r} 
\frac {|\tilde v_r|}{v^\prime_z}
\label{e:diffusn_condn}
\end{equation}
As in the treatment of the $P_{\phi z}$ term above, typical AGN values
$\bar \rho_0/\bar \rho^+ \sim 10^6$, $h_{\rm av}/r \sim 10^{-3}$ this condition relies on the ratio of the radial inflow velocity to the turbulent buoyant velocity. Nevertheless, the impression from (\ref{e:diffusn_condn}) is that the Alfven speed in the corona should be fairly high with respect to the local Keplerian speed for turbulent diffusion to be important.

\subsection{Relative importance of Poynting flux components}

We have seen above that the importance of the various terms in the Poynting flux compared to the accretion power generally depends upon the ratio of the coronal Alfv\'{e}n speed to the local Keplerian speed and its relationship to disk and wind parameters. The above relationships give us a good idea of the importance of the various components of the Poynting flux in absolute terms.  Further insight is obtained if we compare the various terms with respect to one another; this informs us of the conditions under which each component is likely to dominate. In these comparisons we remind the reader that $S_z^{(1)}$ is associated with the power $P_{\phi z}$, $S_z^{(2)}$ is associated with wind-advective part of the Poynting flux and $S_z^{(3)}$ is associated with the turbulent diffusive part.

The ratio 
\begin{equation}
\frac {S_z^{(2)}}{S_z^{(1)}} \simeq
\frac {\langle B^{\prime 2}\rangle^+}{-\langle B^\prime_\phi B^\prime_z  \rangle^+} \, 
\frac {\tilde v_z}{\vK} 
\sim \frac {\langle B^{\prime 2}\rangle^+}{-\langle B^\prime_\phi B^\prime_z  \rangle^+} \,
\frac {h_{\rm av}}{r} \, \frac {\tilde v_z}{c_0}
\end{equation}
Since $\tilde v_z \ltapprox c_0$, the vertical magnetic field would have to satisfy
$B^\prime_z/B^\prime \ltapprox h_{\rm av}/r \ll 1$ in order that the wind advective power dominate $P_{\phi z}$.

Similarly, the ratio of the turbulent diffusive power to the $P_{\phi z}$ is determined by the ratio
\begin{equation}
\frac {S_z^{(3)}}{S_z^{(1)}} \sim
\frac {\langle \case {B^{\prime 2}}{4 \pi}  v^\prime_z\rangle^+}
{-\langle \case{B^\prime_\phi B^\prime_z}{4 \pi} \rangle^+ \vK} 
\sim  \frac {\langle B^{\prime 2}\rangle^+}{-\langle B^\prime_\phi B^\prime_z  \rangle^+} \,
\frac {h_{\rm av}}{r} \, \frac {v^\prime_z}{c_0}
\end{equation}
Again, given that $v^\prime_z \ltapprox c_0$ the vertical component of magnetic field would have to be extremely small for the turbulent diffusive term to dominate.
 
\subsection{The limiting case of vertical transport only}

In a real disk, angular momentum is probably transported by a combination of both radial and vertical fluxes. The limiting case of radial transport only has, of course, been well--explored in the past.
Let us now investigate the opposite limiting case of vertical transport only, specifically as it affects the estimate of the power of the disk wind.
In this limiting case, equation~(\ref{e:angmom}) becomes:
\begin{equation}
\dr \left[ \Mdota \tilde v_\phi r \right]
=\tilde v_\phi r \frac{{\rm d}\Mdota}{{\rm d}r}
- 4\pi r^2 \langle t_{\phi z} \rangle^+   
\label{e:angmom_limit}
\end{equation}
In this limit, the non-kinetic, non-gravitational part of the wind power is dominated by $P_{\phi z}$ so that
\begin{equation}
\Pw \simeq \int_{\ro}^\infty
\left[ - \frac {1}{2} \frac {GM}{r} \frac{{\rm d}\Mdota}{{\rm d}r} 
- 4 \pi r \langle t_{\phi z} \rangle^+ \vK \right] \> {\rm d}r
\label{e:Pw}
\end{equation}
Using equation~(\ref{e:angmom_limit}) the Poynting flux term can be expressed as:
\begin{equation}
- 4 \pi r \langle t_{\phi z} \rangle^+ \vK = \frac {\Mdota \vK}{r} \frac {d}{dr} \left( r \vK \right) 
\end{equation}
Inserting this expression into the integral~(\ref{e:Pw}) for the wind power, one obtains:
\begin{equation}
\Pw \simeq \frac{1}{2} \frac {GM \Mdota(\ro)}{\ro}
\label{e:Pw_limit}
\end{equation}
i.e. the wind power is equal to the accretion power. The corresponding disk luminosity is
\begin{equation}
\Ld \simeq \PQ
\end{equation}
i.e. the entire disk luminosity is attributed to external feedback heating
from the corona. In this limiting case, some of the wind power would be dissipated in the corona and the rest would escape to infinity.

There is a difference between the power in such a wind and the power in a centrifugally driven wind with a nett magnetic flux threading the disc. For example the power in a \citet{blandpayne82} wind is
\begin{equation}
P_{\rm w, BP} \propto B_0^2 (\ro) \ro^2 \left( \frac {GM}{\ro} \right)^{1/2}
\end{equation}
and depends upon the nett magnetic flux in the disk, as well as the Keplerian velocity at the inner radius. 

In the theory we have developed here, the nett flux is zero, so that the magnetic field plays a different role than that envisaged in centrifugally driven flows. Since the $r \phi$ stresses play no role, the result represented in equation~(\ref{e:Pw_limit}) is inevitable. The binding energy released by the accreting disk material has to be manifest in the power of the wind.

\section{Summary and Conclusions}

In this paper, we have established a self-consistent  framework for the 
theory of magnetized, turbulent disk accretion around black holes.
Our formalism is the first to consistently treat turbulent disks using a
robust statistical averaging procedure that explicitly includes
dynamical equations for the evolution of the magnetic field together with the 
conservation equations for mass, momentum and energy transport.
We have paid special attention to  vertical transport of conserved quantities
and consistently related such transport to the dynamical structure of the underlying disk.
Although the nett magnetic flux is assumed to be zero,
the formalism is nonetheless sufficiently general to allow the
straightforward inclusion of a systematic nett mean-field component.

Our main results can be summarized as follows:
\begin{enumerate}
\item We have derived a comprehensive set of equations describing the
transport of  mass, momentum, internal energy, turbulent kinetic and magnetic
energies, and total energy.
The statistically-averaged conservation equations are completely general and
are applicable to other subject areas involving turbulent magnetic 
fields.
\item  We have applied the statistically averaged equations to a geometrically-thin, 
optically-thick accretion disk that is stationary and axisymmetric in the mean.
We have demonstrated that when vertical transport 
of mass, radial, vertical and angular momentum, and energy is self-consistently
treated, the general equations include additional terms related to a disk wind
and turbulent azimuthal--vertical stresses on the disk surface.
\item We have shown that the total azimuthal-vertical stress can have a significant dynamical and energetic effect on the disk even though it may be numerically small compared to the radial--azimuthal stress that has dominated a large amount of accretion disk theory and simulation, to date. Note, however, that the importance of this stress has also been realized in accretion--outlfow models \citep[see][ for a review]{konpud00}.
\item We have derived an expression for the radiative luminosity from the disk 
photosphere and shown clearly how this relates to the mechanical power in a
wind and to the Poynting flux, thereby identifying the possible sources
responsible for powering coronae and/or outflows from accreting black holes. This expression also entails a different distribution of radiative flux than a standard accretion disk. This in turn affects the integrated spectrum. Again, we defer the details to future work.
\item We have discussed the three main sources of Poynting flux into the corona -- a component associated with the product of the  azimuthal - vertical component of the turbulent magnetic stress, a component associated with wind advection of magnetic energy and a component associated with turbulent diffusion of magnetic field from the disk into the corona . The first component probably dominates in most cases even if the azimuthal - vertical stress is quite small in comparison to the radial - azimuthal stress. In the course of the analysis of the condition for a wind and the conditions for a significant Poynting flux into the corona, the ratio of the coronal Alfv\'{e}n emerges as a critical parameter. When this ratio is of order unity, important magnetic effects are clearly present. However, there is also the prospect of significant effects when this parameter is less than unity. This region of parameter space is currently a relatively unexplored avenue of research in black hole accretion disks.
\item In the limiting case, when all of the angular momentum transport is through the vertical-azimuthal stress, we have shown that the wind power, at the base of the wind, is exactly equal to the  accretion power. Some of this power would be dissipated in the corona. This is the first time that a coupled  disk-corona model has demonstrated, in a physically consistent fashion, the possibility of significant power emanating from the corona.
\item This is also the first time that the power of a disk wind has been dynamically linked to the process of accretion. In models with a nett magnetic flux the wind power is related to the strength of the magnetic field.
\item The existence of a coronal wind and the production of intense coronal emission are inextricably linked. The major influence in heating the corona is the stress that is responsible for transporting angular momentum vertically. The wind is essential to transport this angular momentum away from the disk.
\end{enumerate}

In a realistic disk, we expect that both radial--azimuthal and azimuthal--vertical stresses would be
involved in the transport of angular momentum, as well as a nett mass loss from the
innermost regions.
Nevertheless, our limiting solution provides a good physical basis for the commonly held notion that accretion power could be channelled into significant coronal emissivity and/or outflow comprised of electromagnetic and bulk kinetic
components. Thus there is the prospect of explaining not only the coronal emission from radio-quiet
AGN and Galactic Balck Hole Candidates (GBHCs), but also systems such as Ultra-Luminous X-ray (ULX) sources and analogous AGN sources such as Broad Absorption Line (BAL) quasars where large mass outflows (e.g. $\Mdotw \sim \dot M_{\rm Edd}$) are inferred
\citep[see for example][and references therein]{kingpounds03}.

With a self-consistent framework now established, we are in 
the position of being able to consider further, via specific models,
the complex relationship between disk, corona and outflows in a variety of
sources, and to examine more thoroughly the conditions for the initiation of a wind and the
implications for the general  structure of the immediate environment
of accreting black holes.


\begin{thebibliography}{}

\bibitem[Balbus(2003)]{balbus03}
Balbus, S.~A., 2003, \araa, 41, 555

\bibitem[Balbus \& Hawley(1991)]{balbhaw91}
Balbus, S.~A., \& Hawley, J.~F., 1991, \apj, 376, 214

\bibitem[Balbus \& Hawley(1998)]{balbhaw98}
Balbus, S.~A., \& Hawley, J.~F., 1998, Rev. Mod. Phys., 70, 1

\bibitem[\protect\citeauthoryear{Becker, Subramanian \& Kazanas}
{Becker et al.}{2001}]{becker01}
Becker, P. A., Subramanian, P., \& Kazanas, D. 2001, \apj, 552, 209

\bibitem[Begelman(2002)]{begelman02}
Begelman, M. C., 2002, \apj, 568, L97

\bibitem[Bicknell(1984)]{bicknell84}
Bicknell, G.~V. 1984, \apj, 286, 68

\bibitem[Bicknell(1986)]{bicknell86}
Bicknell, G.~V. 1986, \apj, 305, 109

\bibitem[Bisnovatyi-Kogan \& Blinnikov(1977)]{bisblin77}
Bisnovatyi-Kogan, G.~S., \& Blinnikov, S.~I., 1977, \aap, 59, 111

\bibitem[Bisnovatyi-Kogan \& Lovelace(1997)]{bkl97}
Bisnovatyi-Kogan, G.~S., \& Lovelace, R. V. E., 1997, \apj, 486, L43

\bibitem[Bisnovatyi-Kogan \& Lovelace(2000)]{bkl00}
Bisnovatyi-Kogan, G.~S., \& Lovelace, R. V. E., 2000, \apj, 529, 978

\bibitem[Blaes(2002)]{blaes02}
Blaes, O., 2002, in ``Accretion Disks, Jets, and High Energy Phenomena
in Astrophysics'', Proc. of session LXXVIII of Les Houches Summer School,
Chamonix, France, eds. F. Menard, G. Peletier, G. Henri, V. Beskin, and
J. Dalibard (EDP Science: Paris and Springer: Berlin), astro-ph/0211368

\bibitem[Blandford \& Begelman(1999)]{blandbegel99}
Blandford, R.~D., \& Begelman, M. C. 1999, \mnras, 303, L1

\bibitem[Blandford \& Payne(1982)]{blandpayne82}
Blandford, R.~D., \& Payne, D.~G. 1982, \mnras, 199, 883

\bibitem[Bradshaw(1976)]{bradshaw76}
Bradshaw, P. 1976, Turbulence (Berlin: Springer-Verlag)

\bibitem[Brandenburg et~al.(1995)]{brandenburg95}
Brandenburg, A., Nordlund, A., Stein, R. F., \& Torkelsson, U., 1995,
\apj, 446, 741

\bibitem[\protect\citeauthoryear{Cline, Brummell \& Cattaneo}
{Cline et al.}{2003}]{cline03}
Cline, K.~S., Brummell, N.~H., \& Cattaneo, F. 2003, \apj, 588, 630

\bibitem[\protect\citeauthoryear{Di Matteo, Blackman \& Fabian}{Di Matteo
et al.}{1997a}]{dimatt97a}
Di Matteo, T., Blackman, E. G., Fabian, A. C., 1997a, \mnras, 291, L23

\bibitem[\protect\citeauthoryear{Di Matteo, Celotti \& Fabian}{Di Matteo
et al.}{1997b}]{dimatt97b}
Di Matteo, T., Celotti, A., Fabian, A. C., 1997b, \mnras, 291, 805



\bibitem[Favre(1969)]{favre69}
Favre, A. 1969, Statistical Equations of Turbulent gases
(Philadelphia: Society for Industrial and Applied Mathematics), 231



\bibitem[Haardt \&  Maraschi(1991)]{haardt91}
Haardt, F., \& Maraschi, L. 1991, \apj, 380, L51

\bibitem[Haardt \& Maraschi(1993)]{haardt93}
Haardt, F., \& Maraschi, L. 1993, \apj, 413, 507

\bibitem[Haardt et~al.(1994)]{haardt94}
Haardt, F., Maraschi, L., \& Ghisellini, G. 1994, \apj, 432, L95


\bibitem[Hawley \& Balbus(2002)]{hawbalb02}
Hawley, J.~F., \& Balbus, S.~A. 2002, \apj, 573, 738

\bibitem[Hawley \& Krolik(2002)]{hawkrol02}
Hawley, J.~F., \& Krolik, J.~H., 2002, \apj, 566, 164

\bibitem[\protect\citeauthoryear{Hawley, Balbus, \& Stone}
{Hawley et~al.}{2001}]{hawley01}
Hawley, J.~F., Balbus, S.~A., \& Stone, J.~M. 2001, \apj, 554, L52

\bibitem[\protect\citeauthoryear{Hawley, Balbus, \& Winters}
{Hawley et~al.}{1999}]{hawley99}
Hawley, J.~F., Balbus, S.~A., \& Winters, W.~F. 1999, \apj, 518, 394

\bibitem[Heinz \& Begelman(2000)]{heinzbeg00}
Heinz, S., \& Begelman, M. C. 2000, \apj, 535, 104.


\bibitem[Ionson \& Kuperus(1984)]{ionkup84}
Ionson, J. A., \& Kuperus, M. 1984,
\apj, 284, 389

\bibitem[King \& Pounds(2003)]{kingpounds03}
King, A. R., \& Pounds, K. A. 2003,
\mnras, 345, 657

\bibitem[K\"{o}nigl \& Pudritz(2000)]{konpud00}
K\"{o}nigl, A., \& Pudritz, R. E. 2000,
in Protostars and Planets IV,
eds. V. Mannings, A. P. Boss, \& S.~S. Russell, 759

\bibitem[Krause \& Radler(1980)]{krause80}
Krause, F., \& Radler, K.-H. 1980, Mean field Magnetohydrodynamics
and Dynamo Theory (Oxford: Oxford Univ. Press)

\bibitem[\protect\citeauthoryear{Kuncic, Celotti \& Rees}
{Kuncic et al.}{1997}]{kuncic97}
Kuncic, Z., Celotti, A., \& Rees, M.~J. 1997, \mnras, 284, 717

\bibitem[Lazarian \& Vishniac(1999)]{lazvish99}
Lazarian, A., \& Vishniac, E.~T. 1999, \apj, 517, 700

\bibitem[Liang \& Price(1977)]{liangprice77}
Liang, E.~P., \& Price, R.~H. 1977, \apj, 218, 247


\bibitem[\protect\citeauthoryear{Li, Chiueh \& Begelman}
{Li et al.}{1992}]{li92}
Li, Z. Y., Ciueh, T., \& Begelman, M. C. 1992, \apj, 394, 459

\bibitem[\protect\citeauthoryear{Liu, Mineshige \& Shibata}{Liu
et al.}{2002}]{liu02}
Liu, B. F., Mineshige, S., Shibata, K., 2002, \apj, 572, L173

\bibitem[Ustyogova et~al.(2000)]{ustyugova00}
Ustyugova, G. V., Lovelace, R.~V.~E., Romanova, M. M., Li, H., \& Colgate, S. A.
2000,
\apj, 541, 21

\bibitem[\protect\citeauthoryear{Lovelace, Berk, \& Contopolous}{Lovelace et~al.}{1991}]
{lovelace91}
Lovelace, R.~V.~E., Berk, H.~L., \& Contopolous, J. 1991, \apj, 379, 696

\bibitem[\protect\citeauthoryear{Lynden-Bell \& Pringle}{1974}]{lbpring74}
Lynden-Bell, D., \& Pringle, J. E., 1974, \mnras, 168, 603

\bibitem[\protect\citeauthoryear{Meier et~al.}{1997}]{meier97} 
Meier D.L., Edgington S., Godon P., Payne D.G., \& Lind K.R., 1997, Nature, 388, 350

\bibitem[\protect\citeauthoryear{Meier}{1999}]{meier99}
Meier, D.L., 1999, \apj, 522, 753 

\bibitem[\protect\citeauthoryear{Melrose \& McPhedran}{1991}]{melrose91} 
Melrose D.B. \& McPhedran R.C., 1991, 
{\it  Electromagnetic Processes in Dispersive Media},
(Cambridge: Cambridge University Press), 5

\bibitem[\protect\citeauthoryear{Merloni}{2003}]{merloni03}
Merloni, A., 2003, \mnras, 341, 1051

\bibitem[\protect\citeauthoryear{Merloni \& Fabian}{2001a}]{merlfab01a}
Merloni, A., Fabian, A. C., 2001a, \mnras, 321, 549

\bibitem[\protect\citeauthoryear{Merloni \& Fabian}{2001b}]{merlfab01b}
Merloni, A., Fabian, A. C., 2001b, \mnras, 328, 958

\bibitem[\protect\citeauthoryear{Merloni \& Fabian}{2002}]{merlfab02}
Merloni, A., Fabian, A. C., 2002, \mnras, 332, 165

\bibitem[Miller \& Stone(2000)]{millstone00}
Miller, K.~A., \& Stone, J.~M. 2000, \apj, 534, 398

\bibitem[Novikov \& Thorne(1973)]{novthorn73}
Novikov, I.~D., \& Thorne, K.~S. 1973, in Black Holes,
eds. C. DeWitt, \& B. DeWitt (New York: Gordon Breach)

\bibitem[Paczynski(1978)]{pacz78}
Paczynski, B. 1978, Acta. Astr., 28, 241


\bibitem[Parker(1955)]{parker55}
Parker, E.~N. 1955, \apj, 121, 491

\bibitem[Parker(1979)]{parker79}
Parker, E.~N. 1979, Cosmical Magnetic Fields (Oxford: Oxford Univ. Press),
314

\bibitem[Pringle(1981)]{pringle81}
Pringle, J.~E. 1981, \araa, 19, 137

\bibitem[Pringle \& Rees(1972)]{pringrees72}
Pringle, J.~E., \& Rees, M.~J. 1972, \aap, 21, 1

\bibitem[Pudritz \& Norman(1983)]{pudritz83}
Pudritz, R. E., \& Norman, C. A. 1983, \apj, 274, 677

\bibitem[Shakura \& Sunyaev(1973)]{SS73}
Shakura, N.~I., \& Sunyaev, R.~A. 1973, \aap, 24, 337




\bibitem[Stone \& Pringle(2001)]{stonepring01}
Stone, J.~M., \& Pringle, J.~E. 2001, \mnras, 322, 461


\bibitem[Stone et~al.(1996)]{stone96}
Stone, J.~M., Hawley, J.~F., Gammie, C.~F., \&
Balbus, S.~A. 1996, \apj, 463, 656

\bibitem[Svensson \& Zdziarski(1994)]{svensson94}
Svensson, R., \& Zdziarski, A.~A. 1994, \apj, 436, 599




\bibitem[\protect\citeauthoryear{Wardle \& K\"{o}nigl}{1993}]{wardkon93}
Wardle, M., \& K\"{o}nigl, A., 1993, \apj, 410, 218


\end{thebibliography}

\begin{flushright}
\date{\today}
\end{flushright}

\end{document}